\documentclass[twocolumn,showpacs,showkeys,nofootinbib,preprintnumbers,prd]{revtex4-1}

\usepackage{amsmath}
\usepackage{amsfonts}
\usepackage{amssymb}
\usepackage{graphicx}
\usepackage[titletoc]{appendix}
\usepackage{color}
\usepackage{hyperref}
\usepackage{cleveref}
\usepackage[rightcaption]{sidecap}

\usepackage{ragged2e}
\usepackage{caption}
\DeclareCaptionJustification{justified}{\justifying}
\usepackage{subcaption}
\usepackage{comment}

\usepackage{mathtools}

\usepackage{dcolumn}

\usepackage{array}
\usepackage{ctable}
\usepackage{multirow}
\usepackage{siunitx}
\usepackage{tabularx}
\usepackage{booktabs}
\usepackage{supertabular}
\usepackage{verbatim}

\usepackage{cancel}
\usepackage{graphicx}
\usepackage{dcolumn}
\usepackage{bm}
\usepackage{braket}

\begin{document}


\title{Inflationary magnetogenesis in Proca electrodynamics with Schwinger particle production}

\author{Mattia Dubbini}
\email{mattia.dubbini@unicam.it}
\affiliation{Universit\`a di Camerino, Via Madonna delle Carceri, Camerino, 62032, Italy.}
\affiliation{Istituto Nazionale di Fisica Nucleare (INFN), Sezione di Perugia, Perugia, 06123, Italy.}

\author{Orlando Luongo}
\email{orlando.luongo@unicam.it}
\affiliation{Universit\`a di Camerino, Via Madonna delle Carceri, Camerino, 62032, Italy.}
\affiliation{Department of Nanoscale Science and Engineering, University at Albany-SUNY, Albany, New York 12222, USA.}
\affiliation{Istituto Nazionale di Astrofisica (INAF), Osservatorio Astronomico di Brera, 20121 Milano, Italy.}
\affiliation{Istituto Nazionale di Fisica Nucleare (INFN), Sezione di Perugia, Perugia, 06123, Italy.}
\affiliation{Al-Farabi Kazakh National University, Al-Farabi av. 71, 050040 Almaty, Kazakhstan.}

\date{\today}

\begin{abstract}

We investigate inflationary magnetogenesis with massive vector fields, including charged scalar particle production via the Schwinger effect and its backreaction on the background electric energy density. Initially neglecting the Schwinger effect, we quantize the vector field and solve the equations of motion for transverse and longitudinal modes, computing the energy momentum tensor. We then incorporate the backreaction of the electric energy density and show that the longitudinal energy density is subleading compared to the electric background. Since the electric energy density dominates the magnetic contribution, we can treat the former separately from the perturbative contributions. We adopt this approach when studying particle production via the Schwinger effect, assuming that its backreaction acts only on the background electric energy density. The latter is computed using a classical background electromagnetic field minimally coupled to a quantum complex scalar field. The magnetic and longitudinal components are instead described through quantum fields. In this framework, the Schwinger production affects only the electric energy density, as long as it remains dominant compared to the other components. As a consistency check, we verify that for a null coupling constant we recover the same electric energy density as obtained within the fully quantum electromagnetic description. Finally, we show that for non-null coupling constants Schwinger particle production drastically reduces the electric energy density during the final stages of inflation.

\end{abstract}


\maketitle
\tableofcontents

\section{Introduction}{\label{intro}}

The standard Big Bang model provides the established framework for describing the cosmological dynamics of our universe. Despite its success, several theoretical and observational questions remain open, including the fundamental origin of the primordial magnetic fields \cite{Kronberg:1993vk, Grasso:2000wj, Widrow:2002ud, Giovannini:2003yn, Kandus:2010nw, Durrer:2013pga, Subramanian:2015lua, Giovannini:2017rbc} and especially of those detected in cosmic voids through gamma ray observations of distant blazars \cite{Neronov_2010, Tavecchio_2010, Taylor_2011, Dermer_2011, Caprini:2015gga}. Together with constraints from the cosmic microwave background \cite{Planck:2015zrl, Sutton:2017jgr, Jedamzik:2018itu, Paoletti:2018uic} and ultrahigh energy cosmic rays \cite{Bray:2018ipq}, the current magnitude of these fields is constrained to be $10^{-18} \le B_0 \le 10^{-9}\text{ G}$.

The generation of cosmological magnetic fields, commonly referred to as \emph{magnetogenesis}, can be realized through various theoretical schemes \cite{Green:2015fss, Hamada:2025ooy, Sharma:2021rot, Shtanov:2020gjp, Ganz:2025lya, Haque:2020bip}. In particular, several recent approaches \cite{Balaji:2024rvo, Nair:2025iia, Pal:2023vti, Olea-Romacho:2023rhh, Elahi:2021pug} place magnetic field generation in the early universe \cite{Durrer:2022cja, Frion:2020bxc}, particularly during inflation \cite{Adshead:2016iae, Brandenburg:2024awd, Li:2025crs, Durrer:2022emo, Talebian:2020drj}. In these scenarios, magnetic fields can be generated either through nonminimal electromagnetic theories \cite{Savchenko:2018pdr, Sobol:2021jag} or during phase transitions \cite{Domcke:2025itc, Boyanovsky:2002wa, ArteagaTupia:2025awh, Kharzeev:2019rsy} in the early universe.

Another known mechanism consists in breaking the conformal invariance of the Maxwell Lagrangian \cite{Ratra:1991bn, Martin:2007ue, Sobol:2018djj, Luongo:2026zfq, Bamba:2020qdj}, thereby amplifying the production of electric and magnetic fields during inflation. This mechanism, known as inflationary magnetogenesis \cite{Ferreira:2013sqa, Atkins:2025pvg, Maity:2021qps, Nandi:2021lpf}, provides a compelling framework for explaining the generation of magnetic fields in the early universe. In this framework, conformal invariance of the Maxwell Lagrangian is broken by introducing a kinetic coupling between the electromagnetic field and the inflaton \cite{Tripathy:2021sfb}. This coupling is controlled by a function of the inflaton field that decreases throughout inflation and approaches unity at its end, thereby restoring standard electromagnetism. When conformal invariance is broken, quantum fluctuations of the electromagnetic field are amplified by the cosmological expansion and can evolve into a classical electromagnetic background. In homogeneous and isotropic conformally flat spacetimes, it is known that the magnetic energy density is subleading with respect to its electric counterpart \cite{Martin:2007ue}. Previous studies \cite{Sobol:2018djj} show that the latter undergoes exponential growth during inflation until it becomes comparable to the slow roll parameter times the total energy density of the universe. At this stage, backreaction on the inflaton dynamics occurs \cite{Durrer:2023rhc}, and the electric energy density subsequently remains approximately constant while continuing to dominate over the magnetic one.

However, large scale cosmological electric fields are not observed at present, whereas magnetic fields are observationally established. To explain the suppression of electric fields produced during inflation, the Schwinger effect is usually invoked \cite{Bastero-Gil:2025nfi, Bastero-Gil:2023mxm, Kamarpour:2023mhw, Sobol:2019xls, Shakeri:2019mnt, Stahl:2018idd, Kamarpour:2023hoa}, predicting a strong reduction of the electric energy density in the final stages of inflation. The effect consists of the production of particle pairs from the vacuum in the presence of strong electric fields, and within the framework of inflationary magnetogenesis it provides a natural mechanism for terminating the exponential growth of the electric energy density. Indeed, as the electric field grows, particle pairs can be produced through the Schwinger mechanism, and their increasing energy density leads to a strong suppression of the electric field.

So far, magnetogenesis with the Schwinger effect has been explored within the massless scenario \cite{Sobol:2018djj}, for which a time evolution equation for the classical electric energy density can be derived. In such a framework, the Schwinger effect is incorporated by adding a source term proportional to the electric field multiplied by the Schwinger current, either scalar or fermionic \cite{Hayashinaka:2016qqn, Bavarsad:2016cxh}. However, this approach cannot be directly extended to a massive electromagnetic field because an analogous closed evolution equation for the electric energy density is not available.

Inflationary magnetogenesis with massive vector fields has been studied considering a dynamical mass term provided by a coupling with the inflaton field \cite{Lysenko:2025sdo}. The authors showed that, for such a dynamical photon mass, the longitudinal modes emerging as a consequence of gauge invariance breaking yield divergent contributions to the electromagnetic energy density, becoming nonperturbative and even larger than the electric background. Therefore, the backreaction of longitudinal modes needs to be accounted for together with that of the electric energy density. Accordingly, the Schwinger production would affect the longitudinal component of the energy density in addition to the electric one. Such a treatment becomes considerably more involved, and the Schwinger effect in the presence of massive electromagnetic fields has not yet been studied within the framework of inflationary magnetogenesis.

Motivated by the above considerations, in this work we study inflationary magnetogenesis with massive electromagnetic fields, also including the Schwinger effect. We first neglect the Schwinger effect in order to determine the extent to which the longitudinal contributions affect the electromagnetic energy density, and in particular whether they remain perturbative with respect to the electric background, as does the magnetic contribution, or become comparable to it. To this end, we fully quantize the electromagnetic field, solve the mode dynamics, and then compute each term in the electromagnetic energy density as the vacuum expectation value (VEV) of the respective $00$ component of the energy momentum tensor. We show that, considering photon masses within the experimental constraints, the transverse mode equations are not significantly affected relative to the massless scenario, leaving the electric and magnetic energy densities almost unaltered. Overall, the magnetic energy density remains much smaller than its electric counterpart as a consequence of the homogeneity and isotropy of spacetime. In particular, we obtain a present magnetic field amplitude of $B_0\simeq 2.67\times 10^{-17}\text{ G}$, fully consistent with current observational constraints. In addition, we show that longitudinal contributions are comparable to the magnetic energy density and remain perturbative with respect to the electric background. Accordingly, as in the massless case, the electric energy density constitutes the background, whereas the magnetic and longitudinal contributions behave as perturbations.

We then use this result to incorporate the Schwinger effect into our framework. The separation between the electric background and the perturbative contributions allows the two sectors to be treated independently. Since Schwinger production is a nonperturbative effect, it significantly affects only the dominant component of the electromagnetic energy density, which in our framework is the electric component, leaving the perturbative contributions unchanged. In particular, we describe the electric background using a classical electromagnetic field and determine how Schwinger production affects it. The perturbative contributions to the total electromagnetic energy density, namely the magnetic and longitudinal components, remain unaffected by Schwinger production and continue to be described in terms of quantum electromagnetic fields. Our semiclassical approach remains valid as long as the electric background dominates over the other components.

From a Lagrangian viewpoint, we account for the Schwinger effect by introducing a minimal coupling between the massive electromagnetic field and a complex scalar field. The resulting coupled equations are solved considering a classical electromagnetic field and a quantum complex scalar field. The solutions are used to describe both the time evolution of the electric energy density, constituting the background, and the energy density of the produced charged scalar particles. The classical electric energy density is drastically suppressed by the Schwinger production of charged scalar particles, whose energy density grows exponentially during the final stages of inflation.

The paper is organized as follows. In Sect. \ref{overview}, we present a brief overview of the massless framework of inflationary magnetogenesis, discussing the main approaches and results. In Sect. \ref{core}, we study inflationary magnetogenesis with massive vector fields, initially neglecting the Schwinger effect and considering a fully quantum electromagnetic field. We solve the equations of motion for both transverse and longitudinal modes, obtaining analytical solutions. In Sect. \ref{energy_density}, we compute the electric, magnetic and longitudinal components of the electromagnetic energy density, showing that the former constitutes the background while the latter remain perturbative. We then calculate the value of the magnetic field at our time. In Sect. \ref{Schwinger}, we include the Schwinger effect in the treatment. We study the effect of Schwinger production on the classical electric background, while keeping the longitudinal and magnetic components unaltered, and obtain a drastic suppression of electric fields. Finally, in Sect. \ref{conclusions}, we discuss the conclusions and perspectives of our work.

\section{Standard inflationary magnetogenesis}{\label{overview}}

The mechanism of inflationary magnetogenesis consists in breaking the conformal invariance of the Maxwell Lagrangian through a kinetic coupling with the inflaton. The effective Lagrangian density reads
\begin{equation}
\begin{split}
\mathcal{L}_0&=\sqrt{-g}\bigg{[}\frac{1}{2}\partial_{\mu}\phi\partial^{\mu}\phi-V(\phi)-\frac{f^2(\phi)}{4}F^{\mu\nu}F_{\mu\nu}-\frac{M^2_{\text{Pl}}}{16\pi}R\bigg{]},
\end{split}
\label{eq_LagrangianmodelINTRO}
\end{equation}
where the function $f(\phi)$ uniquely determines the kinetic coupling between the inflaton and the electromagnetic field. The gravitational background is described by a spatially flat Friedmann-Lemaître-Robertson-Walker (FLRW) metric,
\begin{equation}
ds^2=dt^2-a^2(t)d{\bf x}^2=a^2(\eta)\left(d\eta^2-d{\bf x}^2\right),
\label{eq_FLRWmetric}
\end{equation}
where $t$ and $\eta$ denote the cosmic and the conformal time, respectively. In particular, the simplest cosmological model for describing the inflationary stage is a de-Sitter universe, where the scale factor is given by
\begin{equation}
a(\eta)=-\frac{1}{H_I\eta},
\label{eq_scalefactor}
\end{equation}
with $H_I$ the (constant) Hubble parameter during inflation. In this work we will consider $\frac{H_I}{M_{\text{Pl}}}\simeq 4.99\times 10^{-6}$, according to Planck data.

Eq. (\ref{eq_scalefactor}) is valid as long as the energy density of the universe is dominated by the inflaton field, that is, during the slow roll phase, when $H(\eta)=H_I$. In particular, concerning the Lagrangian model in Eq. (\ref{eq_LagrangianmodelINTRO}), the de-Sitter approximation holds as long as the sum of the electric and magnetic energy densities is negligible with respect to the inflaton energy density, namely
\begin{equation}
\rho_{\text{E}}+\rho_{\text{B}}\ll \rho_{\text{inf}}\simeq \rho_{\text{bkg}}=\frac{3M^2_{\text{Pl}}H_I^2}{8\pi}.
\label{eq_deSittervalidity}
\end{equation}
Due to the homogeneity and isotropy of the spacetime, the dominant term among the electric and magnetic components is the electric one \cite{Sobol:2018djj, Martin:2007ue}. We remark that, for massless vector fields, the contribution to the energy density coming from the longitudinal modes is null, as these modes do not propagate because of gauge symmetry. Accordingly, previous works considering massless vector fields did not mention such additional terms.

About the coupling function $f$, a simple choice consists in taking
\begin{equation}
f(\eta)=\left(\frac{a}{a_e}\right)^{\alpha},
\label{eq_couplingfunction}
\end{equation}
with $\alpha<0$ to guarantee a decreasing behavior from the beginning to the end of inflation. In fact, $\alpha<-2$ should be taken in order to produce amplified electric and magnetic fields, while $\alpha=-2$ yields constant electric and magnetic energy densities.

However, in principle, $f$ should be a function of the inflaton field $\phi$. Therefore, writing $f(\phi)$ as an explicit function of the conformal time, $f(\eta)$, requires the exact knowledge of the time dependent expression for the inflaton field, that is, $\phi(\eta)$. In this respect, previous works \cite{Sobol:2018djj} have proved that, as long as the produced electromagnetic fields do not affect the inflaton dynamics, considering a power function $f(\eta)$ as in Eq. (\ref{eq_couplingfunction}) corresponds to take
\begin{equation}
f(\phi)=\exp\left\{\frac{3\alpha}{4}\left[1+\sqrt{\frac{2}{3}}\frac{\phi}{M_{\text{Pl}}}-e^{-\sqrt{\frac{2}{3}}\frac{\phi}{M_{\text{Pl}}}}\right]\right\}.
\label{eq_f(phi)}
\end{equation}
In particular, this one-to-one correspondence between Eqs. (\ref{eq_couplingfunction}) and (\ref{eq_f(phi)}) holds in the slow-roll regime, provided that $\phi(\eta)$ is the solution of 
\begin{equation}
\ddot{\phi}+3H\dot{\phi}+V'(\phi)=0,
\label{eq_usualInflatonDyn}
\end{equation}
with $V$ the Starobinsky potential. In general, in inflationary magnetogenesis, the full inflaton equation is given by
\begin{equation}
\ddot{\phi}+3H\dot{\phi}+V'(\phi)=2\frac{\partial(\ln f)}{\partial \phi}\rho_{\text{E}},
\label{eq_inflatonEOM}
\end{equation}
and therefore the equivalence between Eqs. (\ref{eq_couplingfunction}) and (\ref{eq_f(phi)}) is guaranteed as long as the electromagnetic term on the right-hand side is negligible. In this regard, it has been proved that the term on the right-hand side becomes relevant when $\rho_{\text{E}}\simeq \epsilon\rho_{\text{bkg}}$, with $\epsilon$ the slow roll parameter \cite{Sobol:2018djj}. Then, the backreaction of the electric fields becomes relevant in the inflaton dynamics, yielding a different time dependence for $\phi$, and, accordingly, for $f$.
\begin{itemize}
\item[-] This issue is known as the backreaction problem, and can be naturally addressed by solving numerically the coupled dynamics between the inflaton and the electromagnetic field, considering explicitly $f(\phi)$ as in Eq. (\ref{eq_f(phi)}). The numerical output for the electric energy density is a very weakly increasing curve, which can be considered constant compared to the rapid growth undergone until that moment.
\item[-] In fact, it has been shown that, during the backreaction regime, the coupling function $f$ can still be written as a power function, but with power $\alpha=-2$. The resulting electric energy density remains constant during the backreaction regime. This result does not deviate significantly from what one obtains numerically, but has the great advantage of preserving analytical tractability.
\end{itemize}

Even though addressing the backreaction problem is fundamental for the model to be reliable, the main target in inflationary magnetogenesis remains computing the magnetic and electric energy densities. For this purpose, in literature there are basically two approaches.
\begin{itemize}
\item[-] The first approach \cite{Martin:2007ue} consists in quantizing the electromagnetic field and leaving the inflaton classical. In this framework, electric and magnetic energy densities are computed from the respective terms in the 00 component of the electromagnetic energy momentum tensor. Due to the homogeneity and the isotropy of the universe, the electric component turns out to be dominant compared to the magnetic part. Therefore, the electric energy density constitutes the background, whereas the magnetic contribution behaves as a perturbation.
\item[-] The second approach \cite{Sobol:2018djj} is deeply based on the above result. There, indeed, the electric and magnetic sectors are treated independently, and this is possible because of the separation between the electric background and the perturbative magnetic contribution. Concerning the electric sector, starting by classical electromagnetic fields it is possible to find a classical time evolution equation for the electric energy density. On the other hand, the magnetic component is computed perturbatively, described by inhomogeneous quantum fluctuations of the classical electromagnetic field. 

Within this framework, it is also possible to study how the particle production via the Schwinger effect may affect the amplification of the electric fields. In particular, this is done by adding a source term proportional to $E\cdot j$ -- with $j$ the Schwinger current -- in the time evolution equation for the electric energy density.
\end{itemize}

\section{Inflationary magnetogensis within Proca electromagnetism}{\label{core}}

In this work, we study inflationary magnetogenesis in presence of massive electromagnetic fields. Initially, we neglect the Schwinger effect, in order to evaluate
\begin{itemize}
\item[-] the extent to which the mass of the vector fields affects the transverse mode dynamics, and
\item[-] the contribution of the additional longitudinal modes.
\end{itemize}
Afterwards, in the second part of this work, we introduce the Schwinger effect in the model, and investigate the impact of the Schwinger particle production at the background level.

Neglecting the Schwinger effect, we consider the following Lagrangian model,
\begin{equation}
\begin{split}
\mathcal{L}_{\text{m}}&=\sqrt{-g}\bigg{[}-\frac{M^2_{\text{Pl}}}{16\pi}R+\frac{1}{2}\partial_{\mu}\phi\partial^{\mu}\phi-V(\phi)+\\&-\frac{f^2(\phi)}{4}F^{\mu\nu}F_{\mu\nu}+\frac{m^2}{2}A_{\mu}A^{\mu}\bigg{]},
\end{split}
\label{eq_Lagrangianmodel}
\end{equation}
where we have the Proca mass term in addition to $\mathcal{L}_0$.

By virtue of what we discussed in Sect. \ref{overview}, we deal with the backreaction problem analytically. That is, formally we consider the function $f(\phi)$ of Eq. (\ref{eq_f(phi)}), but in fact we consider its time dependent expression $f(\eta)$ of Eq. (\ref{eq_couplingfunction}). We remark that, to account for the backreaction of the electric fields, the power $\alpha$ should be changed when we enter the backreaction regime, namely, when the electric energy density becomes comparable to $\epsilon\rho_{\text{bkg}}$.

In principle, in order to obtain continuous and differentiable solutions to the equations of motion for the fields, we should change $\alpha$ continuously during the transition between the two regimes -- before and after backreaction is established. In this respect, we may assume a time dependent $\alpha(\eta)$, that changes with continuity between the two values $\alpha_{\text{in}}<-2$ and $\alpha_{\text{fin}}=-2$. Even more simply, we can approximate this change as instantaneous, thereby matching the respective field solutions by guaranteeing that they are continuous and differentiable at the transition point.

However, the issue is somewhat more subtle. Indeed, if we match the two field solutions in this way, the various components of the electromagnetic energy density turn out to be discontinuous and non-differentiable, because they depend on $\alpha$ explicitly, and not only through the fields. In fact, the relevant quantities are not the fields themselves, but rather the physical observables. In inflationary magnetogenesis, these correspond precisely to the components of the electromagnetic energy density, and therefore we should require their continuity and differentiability at the transition point. The matching between the two field solutions, obtained for $\alpha_{\text{in}}<-2$ and $\alpha_{\text{fin}}=-2$, should be done accordingly. We will follow this procedure throughout this work.

With respect to the massless scenario, the theoretical model in Eq. (\ref{eq_Lagrangianmodel}) requires the following questions to be addressed.
\begin{itemize}
\item[-] We have an additional propagating degree of freedom -- embedded in the longitudinal modes -- that explicitly breaks gauge invariance.
\item[-] We should study if (and eventually the extent to which) the transverse mode dynamics is affected by the mass of the vector fields.
\item[-] We should quantify the impact of the longitudinal modes on the electromagnetic energy density, specifically checking if the longitudinal contributions are comparable to the electric background or behave as perturbations.
\end{itemize}

As a first step, we do not include the Schwinger particle production in the treatment. Remarkably, this is strictly necessary in order to check if the longitudinal modes bring significant contributions to the electric background or introduce only perturbative corrections. Indeed, as the Schwinger effect is a nonperturbative phenomenon, knowing the nature of the background is fundamental in order to understand which components of the electromagnetic energy density will be affected by the Schwinger particle production.

In this regard, previous works about massive inflationary magnetogenesis \cite{Lysenko:2025sdo} considered strong dynamical mass couplings, providing enormous effective masses to the vector fields. In these scenarios, the longitudinal contributions turn out to be comparable to the electric background and even greater. Accordingly, the Schwinger production would affect the longitudinal component of the energy density in addition to the electric one, and the treatment becomes rather intricate.

Conversely, we consider fixed and small masses, within the experimental constraints, and should check if these leave the electric background unaltered -- providing only perturbative corrections -- or still affect it significantly.

For this purpose, we pursue a fully quantum approach. We consider $A_{\mu}=(A_0,-{\bf A})$, and accordingly we have
\begin{subequations}
\begin{align}
&({\bf E})_i=\frac{F_{0i}}{a}\implies {\bf E}=-\frac{\dot{{\bf A}}+\nabla A_0}{a},
\label{eq_elecricfield}\\
&({\bf B})_i=-\frac{1}{2a^2}\epsilon_{ijk}F_{jk}\implies {\bf B}=\frac{\nabla\wedge {\bf A}}{a^2}.
\label{eq_magneticfield}
\end{align}
\end{subequations}
Therefore, the electromagnetic energy density $\rho_{\text{EM}}$ is given by
\begin{equation}
\rho_{\text{EM}}=\frac{f^2}{2}\braket{{\bf E}^2}+\frac{f^2}{2}\braket{{\bf B}^2}+\frac{m^2}{2}\braket{A_0^2}+\frac{m^2}{2a^2}\braket{{\bf A}^2},
\label{eq_rhoEM}
\end{equation}
and we will refer to the first, second, third and fourth contributions as $\rho_{\text{E}}$ (electric), $\rho_{\text{B}}$ (magnetic), $\rho_0$ and $\rho_{\text{L}}$ (longitudinal)\footnote{In fact, these two contributions contain both transverse and longitudinal modes. However, in analogy with the longitudinal mode, these appear if and only if $m\neq 0$, and thus we refer to them as longitudinal contributions.}, respectively. Two further contributions emerge with respect to the massless case, namely, $\rho_0$ and $\rho_{\text{L}}$, as a consequence of the nonzero mass of the vector fields. In addition, as the longitudinal modes become physical, we also need to account for the longitudinal part of the electric field, while the magnetic field remains purely transverse by construction.

\subsection{Electromagnetic equations of motion}

Starting by the Lagrangian model in Eq. (\ref{eq_Lagrangianmodel}), we obtain the following equations for the electromagnetic field:
\begin{equation}
\partial_{\mu}F^{\mu\nu}+\left(3H+2\frac{\dot{f}}{f}\right)F^{0\nu}+\frac{m^2}{f^2}A^{\nu}=0.
\label{eq_EMequation}
\end{equation}
As a massive electromagnetic field has both longitudinal and transverse propagating modes, in order to solve the equations it is convenient to exploit the Helmholtz decomposition of the electromagnetic vector potential:
\begin{equation}
{\bf A}={\bf A_{\perp}}+{\bf A}_{\parallel},\quad \text{with}\quad \begin{cases}
\nabla{\bf A}=\nabla{\bf A_{\parallel}}\\ \nabla\wedge{\bf A}=\nabla\wedge{\bf A_{\perp}}
\end{cases}.
\label{eq_helmoltzdecomposition}
\end{equation}
Accordingly, the Gauss law is written as
\begin{equation}
\nabla\dot{{\bf A_{\parallel}}}=-\left(\nabla^2-\frac{a^2m^2}{f^2}\right)A_0,
\label{eq_Gausslaw}
\end{equation}
while the spatial components of the electromagnetic equations split into
\begin{subequations}
\begin{align}
&\ddot{{\bf A_{\perp}}}-\frac{\nabla^2{\bf A_{\perp}}}{a^2}+\left(H+2\frac{\dot{f}}{f}\right)\dot{\bf A_{\perp}}+\frac{m^2}{f^2}{\bf A_{\perp}}=0,
\label{eq_EMtransverse}\\
&\ddot{{\bf A_{\parallel}}}\!+\!\left(\!H\!+\!2\frac{\dot{f}}{f}\!\right)\!\dot{\bf A_{\parallel}}\!+\!\frac{m^2}{f^2}{\bf A_{\parallel}}=-\nabla\!\left[\!\dot{A_0}\!+\!\left(\!H\!+\!2\frac{\dot{f}}{f}\!\right)\!A_0\!\right],
\label{eq_EMlongitudinal}
\end{align}
\end{subequations}
for the transverse and longitudinal components, respectively. Finally, a fourth equation can be derived by the divergence of the equations of motion:
\begin{equation}
\dot{A_0}+3HA_0=-\frac{\nabla{\bf A_{\parallel}}}{a^2}.
\label{eq_divEOM}
\end{equation}

\subsection{Quantization of the longitudinal modes}

The quantization of the electromagnetic field in inflationary magnetogenesis has been already explored in scientific literature \cite{Martin:2007ue, Lysenko:2025sdo}. In particular, for the temporal and longitudinal components we have
\begin{subequations}
\begin{align}
&\hat{A_0}(t,{\bf x})=\int\frac{d^3k}{(2\pi)^{\frac{3}{2}}fm}\left[\frac{i}{a}\hat{b}_{{\bf k},\text{L}}\mathcal{D}_L(t,k)e^{i{\bf k}\cdot{\bf x}}+\text{h.c.}\right],
\label{eq_tempquantum}\\
&\hat{{\bf A_{\parallel}}}(t,{\bf x})=\int\frac{d^3k}{(2\pi)^{\frac{3}{2}}fm}\left[\frac{{\bf k}}{k}\hat{b}_{{\bf k},\text{L}}\mathcal{A}_L(t,k)e^{i{\bf k}\cdot{\bf x}}+\text{h.c.}\right],
\label{eq_longquantum}
\end{align}
\end{subequations}
with $
\left[\hat{b}_{{\bf k},\text{L}},\hat{b}^{\dag}_{{\bf p},\text{L}}\right]=\delta^{(3)}({\bf k}-{\bf p})$. For the sake of completeness, we also report the expression for the longitudinal electromagnetic field:
\begin{equation}
\begin{split}
\hat{{\bf E}}_{\parallel}&=-\frac{1}{a}\left(\dot{\hat{{\bf A}}}_{\parallel}+\nabla \hat{A_0}\right)=-\frac{1}{a^2}\int\frac{d^3k}{(2\pi)^{\frac{3}{2}}fm}\times \\&\times \left[\frac{{\bf k}}{k}\hat{b}_{{\bf k},\text{L}}\left(\mathcal{A}_{\text{L}}'+\frac{\alpha}{\eta}\mathcal{A}_{\text{L}}-k\mathcal{D}_{\text{L}}\right)e^{i{\bf k}\cdot{\bf x}}+\text{h.c.}\right].
\end{split}
\label{eq_longelectricfield}
\end{equation}
Handling Eqs. (\ref{eq_EMlongitudinal}) and (\ref{eq_divEOM}), we find
\begin{equation}
\begin{cases}
\dot{\mathcal{A}_{\text{L}}}-\frac{\dot{f}}{f}\mathcal{A}_{\text{L}}=\frac{k}{a}\left(1+\frac{a^2m^2}{k^2f^2}\right)\mathcal{D}_{\text{L}} \\ \dot{\mathcal{D}_{\text{L}}}+\left(2H-\frac{\dot{f}}{f}\right)\mathcal{D}_{\text{L}}=-\frac{k}{a}\mathcal{A}_{\text{L}}
\end{cases},
\label{eq_longmodeequations}
\end{equation}
that is a system of coupled first-order differential equations. With respect to conformal time, writing $f$ explicitly, we get
\begin{equation}
\begin{cases}
\mathcal{A}_{\text{L}}'+\frac{\alpha}{\eta}\mathcal{A}_{\text{L}}=k\left(1+\frac{a^2m^2}{k^2f^2}\right)\mathcal{D}_{\text{L}} \\ \mathcal{D}_{\text{L}}'+\frac{\alpha-2}{\eta}\mathcal{D}_{\text{L}}=-k\mathcal{A}_{\text{L}}
\end{cases}.
\label{eq_longmodeequationsCONFORMALtime}
\end{equation}
We remarkably notice that the quantity in round brackets on the right-hand side of the equation for $\mathcal{A}_{\text{L}}$ can be safely approximated to $1$ for masses $m$ within experimental bounds. In particular, this is true for the following inflationary parameters:
\begin{itemize}
\item[-] the pivot scale is taken at $k_*\simeq 3.15\times 10^{-40}\text{ GeV}$;
\item[-] the number of e-folds after which the pivot scale exits the horizon is $N_*=10$;
\item[-] the total number of e-folds is $N_{\text{f}}=60$.
\end{itemize}
These parameters confine the inflationary epoch in $[\eta_{\text{i}},\eta_{\text{f}}]$, with $\eta_{\text{i}}\simeq -6.99\times 10^{43}\text{ GeV}^{-1}$ and $\eta_{\text{f}}\simeq -6.12\times 10^{17}\text{ GeV}^{-1}$. Accordingly, the crossing-horizon scales at the beginning and the end of inflation are $k_{\text{i}}=-\frac{1}{\eta_{\text{i}}}\simeq 1.43\times 10^{-44}\text{ GeV}$ and $k_{\text{f}}=-\frac{1}{\eta_{\text{f}}}\simeq 1.63\times 10^{-18}\text{ GeV}$, respectively. 

In addition, in the following we will assume Bunch-Davies initial conditions for all the free field modes, and therefore both infrared (IR) and ultraviolet (UV) cutoffs should be set.
\begin{itemize}
\item[-] As an IR cutoff, we take $k>k_{\text{i}}$, \emph{de facto} considering only those wavelengths that have not yet entered the horizon when inflation starts.
\item[-] Then, we select a time dependent UV cutoff, given by $k_H(\eta)\equiv a(\eta)H_I=-1/\eta$. In this way, at each time, we include only the modes that have crossed the Hubble horizon, and have thus classicalized. These modes are those that effectively contribute to the classical energy density, that is what we want to compute in this work.
\end{itemize}
We therefore restrict the momentum integration to $k\in[k_i,k_H(\eta)]$, i.e., leaving open the challenge to formulate the same analysis into a self-consistent quantum gravity scenario. Selecting $k\in[k_{\text{i}},k_H(\eta)]$ at time $\eta$, it is immediate to find
\begin{equation}
\frac{a^2m^2}{k^2f^2}<\frac{m^2}{M^2_{\text{Pl}}}\frac{1}{1.91}\times 10^{63},
\label{eq_negligiblemassinEOMs}
\end{equation}
guaranteeing $\frac{a^2m^2}{k^2f^2}\ll 1$ as long as $\frac{m}{M_{\text{Pl}}}\ll 4.37\times 10^{-32}$, that is, for $m\ll 0.53\text{ meV}$. Experimental bounds are much stricter \cite{Goldhaber:2008xy, Yan:2023kdg, Ryutov:2007zz, Lin:2023jaq, Zhang:2016ruz, Bartlett:2021olb}, constraining an eventual nonzero photon mass to be smaller than $10^{-14}$ or even $10^{-18}\text{ eV}$. Therefore, we can safely approximate Eq. (\ref{eq_longmodeequationsCONFORMALtime}) to
\begin{equation}
\begin{cases}
\mathcal{A}_{\text{L}}'+\frac{\alpha}{\eta}\mathcal{A}_{\text{L}}=k\mathcal{D}_{\text{L}} \\ \mathcal{D}_{\text{L}}'+\frac{\alpha-2}{\eta}\mathcal{D}_{\text{L}}=-k\mathcal{A}_{\text{L}}
\end{cases}.
\label{eq_longmodeequationsCONFORMALtimeNOmass}
\end{equation}
Before of solving the system, we notice that Eq. (\ref{eq_longmodeequationsCONFORMALtimeNOmass}) would yield an identically null longitudinal electric field, as the mode combination in Eq. (\ref{eq_longelectricfield}) would vanish. This is expected, as the longitudinal electric field is a physical observable and emerges if and only if $m\neq 0$. Accordingly, totally neglecting the mass in the equations, we should recover the gauge-invariant case, where ${\bf E}_{\parallel}=0$. Nevertheless, we can consider Eq. (\ref{eq_longmodeequationsCONFORMALtime}) in order to write the longitudinal electric field as
\begin{equation}
\hat{{\bf E}}_{\parallel}=-\frac{m}{f^3}\int\frac{d^3k}{(2\pi)^{\frac{3}{2}}} \left[\frac{{\bf k}}{k}\hat{b}_{{\bf k},\text{L}}\frac{\mathcal{D}_{\text{L}}}{k}e^{i{\bf k}\cdot{\bf x}}+\text{h.c.}\right],
\label{eq_longelectricfieldSIMPLIFIED}
\end{equation}
and then use for $\mathcal{D}_{\text{L}}$ the leading-order solution provided by Eq. (\ref{eq_longmodeequationsCONFORMALtimeNOmass}). In particular, we find exact solutions of Eq. (\ref{eq_longmodeequationsCONFORMALtimeNOmass}) setting the following initial conditions \cite{Lysenko:2025sdo} on $\mathcal{A}_{\text{L}}$:
\begin{equation}
\mathcal{A}_{\text{L}}(-k\eta\to +\infty)\simeq \frac{f}{a}\sqrt{\frac{k}{2}}e^{-ik\eta}.
\label{eq_icAL}
\end{equation}
Accordingly, we find
\begin{subequations}
\begin{align}
&\mathcal{A}_{\text{L}}(\eta)=\frac{f}{a}\sqrt{\frac{k}{2}}\left(1-\frac{i}{k\eta}\right)e^{-ik\eta},
\label{eq_ALsolution}\\
&\mathcal{D}_{\text{L}}(\eta)=-i\frac{f}{a}\sqrt{\frac{k}{2}}e^{-ik\eta}.
\label{eq_DLsolution}
\end{align}
\end{subequations}

\subsection{Quantization of the transverse modes}

For the transverse modes, we have
\begin{equation}
\hat{A}_{\perp j}(t,{\bf x})\!=\!\!\int\!\frac{d^3k}{(2\pi)^{\frac{3}{2}}f}\sum_{\lambda=1}^2\left[\epsilon_j^{(\lambda)}(k)\hat{b}_{{\bf k},\lambda}\mathcal{A}_{\lambda}(t,k)e^{i{\bf k}\cdot{\bf x}}+\text{h.c.}\right],
\label{eq_quantumAtransverse}
\end{equation}
with $
\left[\hat{b}_{{\bf k},\lambda_1},\hat{b}^{\dag}_{{\bf p},\lambda_2}\right]=\delta_{\lambda_1\lambda_2}\delta^{(3)}\left({\bf k}-{\bf p}\right)$.
In analogy with the longitudinal component, we report the quantized expression for the transverse electric field:
\begin{equation}
\begin{split}
(\hat{{\bf E}}_{\perp})_j&=-\frac{\dot{\hat{A}}_{\perp j}}{a}=-\frac{\hat{A}_{\perp j}'}{a^2}=-\frac{1}{a^2f}\int\frac{d^3k}{(2\pi)^{\frac{3}{2}}}\times \\&\times \sum_{\lambda=1}^2\left[\epsilon_j^{(\lambda)}(k)\hat{b}_{{\bf k},\lambda}\left(\mathcal{A}_{\lambda}'+\frac{\alpha}{\eta}\mathcal{A}_{\lambda}\right)e^{i{\bf k}\cdot {\bf x}}+\text{h.c.}\right].
\end{split}
\label{eq_transvelectricfield}
\end{equation}
In addition, from the transverse modes we can also derive the magnetic field:
\begin{equation}
\begin{split}
(\hat{{\bf B}})_m&=(\hat{{\bf B}}_{\perp})_m=\frac{\epsilon_{mij}\partial_i\hat{A}_{\perp j}}{a^2}=\frac{1}{a^2f}\int\frac{d^3k}{(2\pi)^{\frac{3}{2}}}\times \\&\times \sum_{\lambda=1}^2\left[i\epsilon_{mij}k_i\epsilon_j^{(\lambda)}(k)\hat{b}_{{\bf k},\lambda}\mathcal{A}_{\lambda}e^{i{\bf k}\cdot{\bf x}}+\text{h.c.}\right].
\end{split}
\label{eq_magneticfield}
\end{equation}
Concerning the dynamics, Eq. (\ref{eq_EMtransverse}) yields
\begin{equation}
\ddot{\mathcal{A}_{\lambda}}+H\dot{\mathcal{A}_{\lambda}}+\left(\frac{k^2}{a^2}+\frac{m^2}{f^2}-\frac{H\dot{f}}{f}-\frac{\ddot{f}}{f}\right)\mathcal{A}_{\lambda}=0,
\label{eq_transvmodeequations}
\end{equation}
that in conformal time becomes
\begin{equation}
\mathcal{A}_{\lambda}''+k^2\left(1+\frac{a^2m^2}{k^2f^2}-\frac{\alpha(\alpha+1)}{k^2\eta^2}\right)\mathcal{A}_{\lambda}=0.
\label{eq_transvmodeequationsCONFORMALtime}
\end{equation}
As for longitudinal modes, here too masses within the experimental bounds bring a negligible contribution to the equation, that can be accordingly simplified to
\begin{equation}
\mathcal{A}_{\lambda}''+\left(k^2-\frac{\alpha(\alpha+1)}{\eta^2}\right)\mathcal{A}_{\lambda}=0.
\label{eq_transvmodeequationsCONFORMALtimeNOmass}
\end{equation}
The most general solution of Eq. (\ref{eq_transvmodeequationsCONFORMALtimeNOmass}) is
\begin{equation}
\mathcal{A}_{\lambda}(\eta)=\sqrt{-\eta}\left[c_1J_{\alpha+\frac{1}{2}}(-k\eta)+c_2Y_{\alpha+\frac{1}{2}}(-k\eta)\right],
\label{eq_GENERALtransvsolution}
\end{equation}
where $c_1$ and $c_2$ are integration constants to be determined, $J_{\alpha+\frac{1}{2}}(x)$ and $Y_{\alpha+\frac{1}{2}}(x)$ are the Bessel functions of first and second kind, respectively, of order $\nu_\alpha=\alpha+\frac{1}{2}$. For this kind of solutions, we can safely assume Bunch-Davies initial conditions, yielding
\begin{equation}
\mathcal{A}_{\lambda}(\eta)=e^{i\frac{\pi}{2}\left(1+\alpha\right)}\sqrt{-\eta}\frac{\sqrt{\pi}}{2}H_{\alpha+\frac{1}{2}}^{(1)}(-k\eta),
\label{eq_transvsolution}
\end{equation}
with $H_{\alpha+\frac{1}{2}}^{(1)}(x)$ the first-kind Hankel function of order $\nu_{\alpha}$.

We notice that, differently from $\mathcal{A}_{\text{L}}$ and $\mathcal{D}_{\text{L}}$, the transverse modes $\mathcal{A}_{\lambda}$ depend explicitly on $\alpha$. Hence, the transverse electric and magnetic fields will depend on $\alpha$ too. Accordingly, all the contributions to the energy density involving ${\bf A}_{\perp}$, ${\bf E}_{\perp}$ and ${\bf B}$ will be affected by the backreaction problem that we discussed at the beginning of this section. 

To address this question, the following prescription should be pursued for each of these components of the energy density. 
\begin{itemize}
\item[-] Using Eq. (\ref{eq_transvsolution}) to compute the energy density until backreaction regime is established, then
\item[-] taking the general solution in Eq. (\ref{eq_GENERALtransvsolution}) with $\alpha=-2$, leaving the integration constants undetermined,
\item[-] computing the corresponding energy density,
\item[-] determining the integration constants requiring continuity and differentiability of the energy density at the transition point, when backreaction ideally instantly starts.
\end{itemize}
We will refer to the solutions valid until the backreaction starts as $\mathcal{A}_{\lambda}^{(\text{in})}(\eta)$, given by Eq. (\ref{eq_transvsolution}). On the other hand, the solutions valid in the backreaction regime can be generally written as
\begin{equation}
\mathcal{A}_{\lambda}^{(\text{fin})}(\eta)=\sqrt{-\eta}\left[c_1J_{-\frac{3}{2}}(-k\eta)+c_2Y_{-\frac{3}{2}}(-k\eta)\right],
\label{eq_transvsolutionAFTERbackreaction}
\end{equation}
where $c_1$ and $c_2$ should be determined through the prescription presented above.

As a last comment, we will see that, even though in principle also ${\bf E}_{\parallel}$ depends on $\alpha$ through $f$, for the related energy density contribution we do not need to construct the curve as discussed above.

\section{Calculation of the electromagnetic energy density}{\label{energy_density}}

In this section, we will compute the various contributions to the electromagnetic energy density, written explicitly in Eq. (\ref{eq_rhoEM}). We start by the longitudinal components, namely, $\rho_0$, $\rho_{\text{L}\parallel}\equiv \frac{m^2}{2a^2}\braket{{\bf A}_{\parallel}^2}$ and $\rho_{\text{E}\parallel}\equiv \frac{f^2}{2}\braket{{\bf E}_{\parallel}^2}$.

In particular, $\rho_0$ and $\rho_{\text{L}\parallel}$ do not depend on $\alpha$, and thus keep the same form for the whole duration of inflation. For them, we find
\begin{subequations}
\begin{align}
&\frac{\rho_0(\theta)}{M^4_{\text{Pl}}}=\frac{1}{32\pi^2}\left(\frac{H_I}{M_{\text{Pl}}}\right)^4\left(1-\theta^4\right),
\label{eq_rho0}\\
&\frac{\rho_{\text{L}\parallel}(\theta)}{M^4_{\text{Pl}}}=\frac{1}{32\pi^2}\left(\frac{H_I}{M_{\text{Pl}}}\right)^4\left(3-2\theta^2-\theta^4\right),
\label{eq_rhoLparallel}
\end{align}
\end{subequations}
where $\theta=-k_{\text{i}}\eta$. It is immediate to check that $\theta=1$ at the beginning of inflation, and as inflation proceeds $\theta$ approaches smaller values, until reaching its minimum $\theta_{\text{min}}\simeq 8.75\times 10^{-27}$ at the end of inflation.

Both $\rho_0$ and $\rho_{\text{L}\parallel}$ grow as inflation proceeds, approaching the following plateau regions in the limit of $\theta\ll 1$:
\begin{subequations}
\begin{align}
&\frac{\rho_0(\theta\ll 1)}{M^4_{\text{Pl}}}=\frac{1}{32\pi^2}\left(\frac{H_I}{M_{\text{Pl}}}\right)^4,
\label{eq_rho0asymptotic}\\
&\frac{\rho_{\text{L}\parallel}(\theta\ll 1)}{M^4_{\text{Pl}}}=\frac{3}{32\pi^2}\left(\frac{H_I}{M_{\text{Pl}}}\right)^4.
\label{eq_rhoLparallelasymptotic}
\end{align}
\end{subequations}
It is straightforward to verify that both these contributions remain considerably lower than the background energy density, defined by Eq. (\ref{eq_deSittervalidity}).

Now, we want to compute $\rho_{\text{E}\parallel}$. In principle, this contribution depends on $\alpha$ through the function $f$, and therefore we should distinguish between the two regimes -- before and after backreaction starts -- following the prescription discussed at the end of Sect. \ref{core}. 

However, we will prove that, for $\rho_{\text{E}\parallel}$, this construction is not necessary, at least for the purposes of this work. Indeed, if we compute the solution valid before backreaction begins, we find
\begin{equation}
\frac{\rho_{\text{E}\parallel}^{(\text{in})}(\theta)}{M^4_{\text{Pl}}}=\frac{1}{16\pi^2}\left(\frac{m}{M_{\text{Pl}}}\right)^2\left(\frac{H_I}{M_{\text{Pl}}}\right)^2\left(\frac{\theta}{\theta_{\text{min}}}\right)^{2\alpha}\left(1-\theta^2\right).
\label{eq_longrhoE}
\end{equation}
As inflation proceeds, and accordingly $\theta$ decreases approaching values much smaller than the unity, $\rho_{\text{E}\parallel}^{(\text{in})}$ grows like $\theta^{2\alpha}$, with $\alpha<0$. Therefore, if $\rho_{\text{E}\parallel}^{(\text{in})}$ kept the same form for the whole duration of the inflation, its maximum would be at $\theta=\theta_{\text{min}}$, for which
\begin{equation}
\frac{\rho_{\text{E}\parallel}^{(\text{in})}(\theta_{\text{min}})}{M^4_{\text{Pl}}}=\frac{1}{16\pi^2}\left(\frac{m}{M_{\text{Pl}}}\right)^2\left(\frac{H_I}{M_{\text{Pl}}}\right)^2.
\label{eq_maximumrhoEparallel}
\end{equation}
For masses $m$ within experimental bounds, the value in Eq. (\ref{eq_maximumrhoEparallel}) is much lower than the plateau regions reached by $\rho_0$ and $\rho_{\text{L}\parallel}$. In fact, we expect the real maximum of $\rho_{\text{E}\parallel}$ to be even lower than that of $\rho_{\text{E}\parallel}^{(\text{in})}$, as the former would not grow like $\theta^{2\alpha}$ (with $\alpha<-2$) but like $\theta^{-4}$, thus even more slowly. Therefore, if we assume continuity and differentiability of $\rho_{\text{E}\parallel}$ at the transition point between the two regimes, after backreaction we expect a flatten curve, and accordingly the real maximum to be lower than that reached ideally by $\rho_{\text{E}\parallel}^{(\text{in})}$. For these reasons, from now on we will neglect the contribution of $\rho_{\text{E}\parallel}$ in our treatment.

Before of plotting and visualizing the relevant energy density contributions coming from the longitudinal modes, we should briefly discuss about the massless limit of the theory. In particular, the expressions for $\rho_0$ and $\rho_{\text{L}\parallel}$ do not depend on $m$, and this is due to the normalization factor $\frac{1}{m}$ in front of $\hat{A}_{\parallel}$ and $\hat{A}_0$. Therefore, one may wonder if in the massless limit we really obtain null $\rho_0$ and $\rho_{\text{L}\parallel}$, as expected for having a continuous massless limit, or not. On the other hand, for $m=0$ the longitudinal electric field is exactly zero -- and the same holds for $\rho_{\text{E}\parallel}$ -- and in this case the massless limit is continuous.

The question can be fully understood looking at the Lagrangian density in Eq. (\ref{eq_Lagrangianmodel}), and is strictly related to the gauge invariance. Indeed, we notice that
\begin{itemize}
\item[-] the gauge-invariant quantity, that is the longitudinal electric field (or the associated energy density), recovers correctly the massless limit, while
\item[-] the non-gauge-invariant quantities, that are $\rho_0$ and $\rho_{\text{L}\parallel}$, do not.
\end{itemize}
Accordingly, we expect that the massless limit may be recovered continuously if and only if gauge invariance is completely restored, e.g. through the Stueckelberg mechanism. However, in this work we do not deal with gauge invariance restoration, reserving this analysis for a future paper.

\begin{figure}
\includegraphics[width=0.5\textwidth]{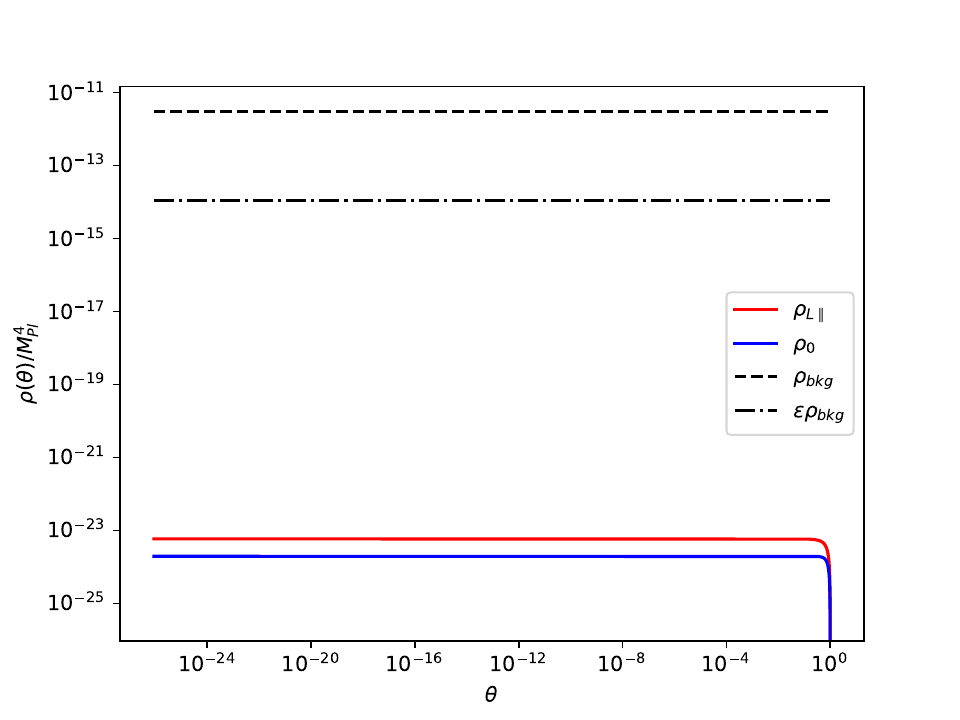}
\caption{Dominant components among the longitudinal contributions to the electromagnetic energy density, say $\rho_0$ (blue line) and $\rho_{\text{L}\parallel}$ (red line), versus the rescaled conformal time $\theta=-k_{\text{i}}\eta$.}
\label{fig_longitudinalcontributions}
\end{figure}

Now, we can visualize the plots. In Fig. \ref{fig_longitudinalcontributions}, we show the two dominant contributions among the longitudinal ones, namely, $\rho_0$ and $\rho_{\text{L}\parallel}$. They are plotted compared to the background energy density, $\rho_{\text{bkg}}$, and $\epsilon\rho_{\text{bkg}}$. We remark that these two energy density contributions are present uniquely because we have a massive electromagnetic field, even though very light.

Then, we compute the transverse contributions. Among the full expression for the electromagnetic energy density in Eq. (\ref{eq_rhoEM}), the transverse components are $\rho_{\text{B}}$, $\rho_{\text{L}\perp}\equiv \frac{m^2}{2a^2}\braket{{\bf A}_{\perp}^2}$ and $\rho_{\text{E}\perp}\equiv \frac{f^2}{2}\braket{{\bf E}_{\perp}^2}$. Depending on the transverse modes, all of these contributions depend explicitly on $\alpha$, and we need to construct them through the prescription illustrated at the end of Sect. \ref{core}.

We start from $\rho_{\text{E}\perp}$. Using the transverse electric field in Eq. (\ref{eq_transvelectricfield}), we find the following general expression for $\rho_{\text{E}\perp}$, valid for each $\alpha$ and independent of the choice of the initial conditions:
\begin{equation}
\rho_{\text{E}\perp}(\eta)=\frac{1}{2\pi^2a^4}\int_{k_i}^{k_H(\eta)} dk k^2\left|\mathcal{A}_{\lambda}'+\frac{\alpha}{\eta}\mathcal{A}_{\lambda}\right|^2.
\label{eq_rhoEperpGENERAL}
\end{equation}
When we specialize Eq. (\ref{eq_rhoEperpGENERAL}) for the solutions in Eq. (\ref{eq_transvsolution}), valid before backreaction, we get
\begin{equation}
\frac{\rho_{\text{E}\perp}^{(\text{in})}(\theta)}{M^4_{\text{Pl}}}=\frac{1}{8\pi}\left(\frac{H_I}{M_{\text{Pl}}}\right)^4\int_{\theta}^1dy y^4\left|H_{\alpha-\frac{1}{2}}^{(1)}(y)\right|^2.
\label{eq_rhoEperpINITIAL}
\end{equation}
Then, Eq. (\ref{eq_rhoEperpGENERAL}) should be recomputed for $\alpha=-2$ and using the transverse modes in Eq. (\ref{eq_transvsolutionAFTERbackreaction}). The resulting curve depends on the integration constants $c_1$ and $c_2$, that are determined by requiring continuity and differentiability at the transition point between the two regimes. In particular, we name this point $\theta_*$, and we determine it assuming
\begin{equation}
\rho_{\text{E}\perp}(\theta_*)=\gamma\epsilon\rho_{\text{bkg}},
\label{eq_thetastar}
\end{equation}
where $\gamma$ is a coefficient smaller than the unity to be decided consistently with numerical results \cite{Sobol:2018djj}. Eq. (\ref{eq_thetastar}), in principle, according to Eq. (\ref{eq_deSittervalidity}), should have the sum of $\rho_{\text{E}}$ and $\rho_{\text{B}}$ on the left-hand side. However, we will prove that $\rho_{\text{E}\perp}$ is dominant with respect to both $\rho_{\text{E}\parallel}$ and $\rho_{\text{B}}$, justifying the simplification used in Eq. (\ref{eq_thetastar}).

The coefficient $\gamma$ should be determined such that the backreaction starts slightly before the electric energy density becomes equal to $\epsilon\rho_{\text{bkg}}$, when it would affect the inflaton dynamics. In particular, we take $\gamma=0.1$ in this work, but different choices would lead to analogous results as long as $\gamma\lesssim 1$. In addition, we can find $\epsilon$ by the value of the power spectrum at the pivot scale, that is $P_{\text{s}}\simeq 2.1\times 10^{-9}$, yielding $\epsilon\simeq 3.76\times 10^{-3}$.

\begin{figure}
\includegraphics[width=0.5\textwidth]{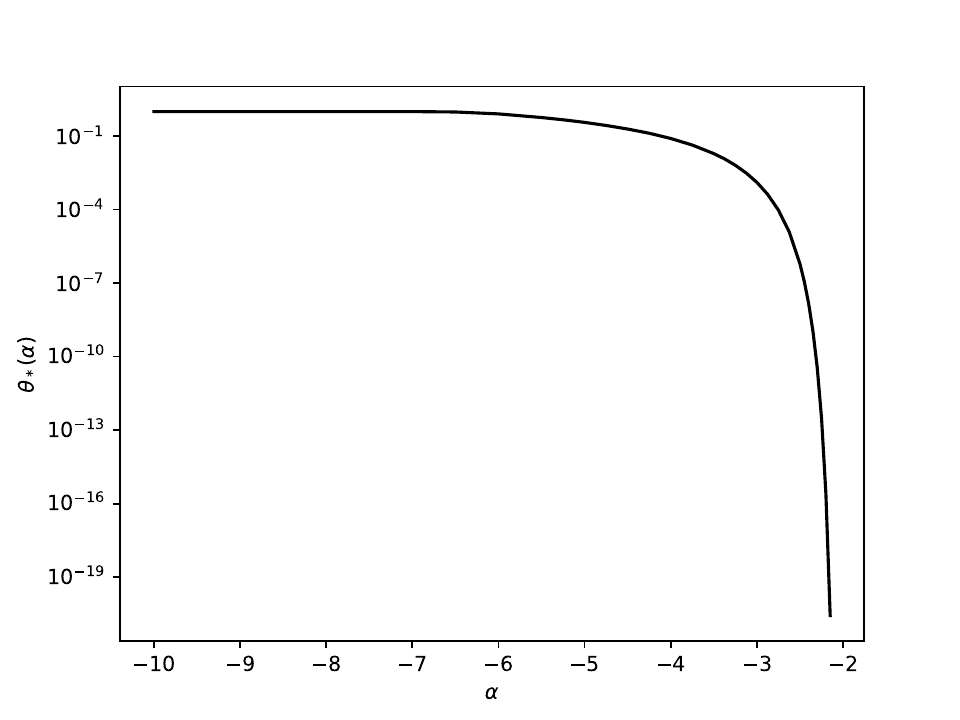}
\caption{Numerical curve $\theta_*(\alpha)$ obtained from Eq. (\ref{eq_thetastarVSalpha}).}
\label{fig_thetastarVSalpha}
\end{figure}

Once we selected a precise value for $\gamma$ and computed $\epsilon$, Eq. (\ref{eq_thetastar}) gives a precise relation among $\theta_*$ and $\alpha$,
\begin{equation}
\int_{\theta_*}^1dy y^4\left|H_{\alpha-\frac{1}{2}}^{(1)}(y)\right|^2=\frac{3\epsilon}{10}\left(\frac{M_{\text{Pl}}}{H_I}\right)^2\simeq 4.53\times 10^7,
\label{eq_thetastarVSalpha}
\end{equation}
from which we numerically obtain the functional dependence $\theta_*(\alpha)$. The resulting curve $\theta_*(\alpha)$ is shown in Fig. \ref{fig_thetastarVSalpha}. In particular, we find that as $\alpha\to -2$ the corresponding $\theta_*$ considerably decreases, and below a certain threshold, in our case almost $\alpha=-2.15$, Eq. (\ref{eq_thetastarVSalpha}) does not admit solutions at all. Conversely, as $\alpha$ increases in modulus, $\theta_*$ approaches the value of $1$, suggesting that the backreaction regime would start immediately at the beginning of inflation. In this work, unless differently specified, we use $\alpha=-2.5$, for which we get $\theta_*(-2.5)\simeq 5.98\times 10^{-7}$.

Once we determined the transition point $\theta_*$ at the varying of $\alpha$ and we evaluated its value for $\alpha=-2.5$, we can extend the expression in Eq. (\ref{eq_rhoEperpINITIAL}) for the whole duration of the inflation. In particular, we take the general expression for $\rho_{\text{E}\perp}$ from Eq. (\ref{eq_rhoEperpGENERAL}) and compute it for $\alpha=-2$, using the general mode solutions in Eq. (\ref{eq_transvsolutionAFTERbackreaction}). We name the resulting function $\rho_{\text{E}\perp}^{(\text{fin})}$, and it depends on $c_1$ and $c_2$. According to our prescription, we find the integration constants requiring continuity and differentiability in $\theta_*$ between $\rho_{\text{E}\perp}^{(\text{in})}$ and $\rho_{\text{E}\perp}^{(\text{fin})}$. In this way, for $\rho_{\text{E}\perp}^{(\text{fin})}$ we get
\begin{equation}
\begin{split}
&\frac{\rho_{\text{E}\perp}^{(\text{fin})}(\theta)}{M^4_{\text{Pl}}}=\frac{\pi}{128}\left(\frac{H_I}{M_{\text{Pl}}}\right)^4\left(\frac{\theta_*}{\theta}\right)^2\times \\&\times \int_{\theta}^1dy y^6\left|e_2\left(\frac{\theta_*}{\theta}y\right)Y_{-\frac{5}{2}}(y)-e_1\left(\frac{\theta_*}{\theta}y\right)J_{-\frac{5}{2}}(y)\right|^2,
\end{split}
\label{eq_rhoEperpAFTER}
\end{equation}
where $e_1(x)$ and $e_2(x)$ are defined in App. \ref{app_usefulfunctions}. Finally, we have
\begin{equation}
\frac{\rho_{\text{E}\perp}(\theta)}{M^4_{\text{Pl}}}=\frac{1}{M^4_{\text{Pl}}}\begin{cases}
\rho_{\text{E}\perp}^{(\text{in})}(\theta) & \theta>\theta_* \\ \rho_{\text{E}\perp}^{(\text{fin})}(\theta) & \theta<\theta_*
\end{cases}.
\label{eq_rhoEperpFULL}
\end{equation}
Then, we compute the magnetic energy density, $\rho_{\text{B}}$. Using the magnetic field in Eq. (\ref{eq_magneticfield}), we find the following general expression for $\rho_{\text{B}}$, valid for each $\alpha$ and independent of the choice of the initial conditions:
\begin{equation}
\rho_{\text{B}}(\eta)=\frac{1}{2\pi^2a^4}\int_{k_i}^{k_H(\eta)}dk k^4|\mathcal{A}_{\lambda}|^2.
\label{eq_rhoBGENERAL}
\end{equation}
When we specialize Eq. (\ref{eq_rhoBGENERAL}) for the solutions in Eq. (\ref{eq_transvsolution}), valid before backreaction, we find
\begin{equation}
\frac{\rho_{\text{B}}^{(\text{in})}(\theta)}{M^4_{\text{Pl}}}=\frac{1}{8\pi}\left(\frac{H_I}{M_{\text{Pl}}}\right)^4\int_{\theta}^1dy y^4\left|H_{\alpha+\frac{1}{2}}^{(1)}(y)\right|^2.
\label{eq_rhoBINITIAL}
\end{equation}
As we did for $\rho_{\text{E}\perp}$, in order to compute the magnetic energy density after backreaction we need to compute Eq. (\ref{eq_rhoBGENERAL}) for $\alpha=-2$, using the general solutions for the transverse modes in Eq. (\ref{eq_transvsolutionAFTERbackreaction}). We name the resulting function $\rho_{\text{B}}^{(\text{fin})}$. Once again, we find the integration constants $c_1$ and $c_2$ following our prescription, yielding
\begin{equation}
\begin{split}
&\frac{\rho_{\text{B}}^{(\text{fin})}(\theta)}{M^4_{\text{Pl}}}=\frac{\pi}{128}\left(\frac{H_I}{M_{\text{Pl}}}\right)^4\left(\frac{\theta_*}{\theta}\right)^2\times \\&\times \int_{\theta}^1dy y^6\left|g_2\left(\frac{\theta_*}{\theta}y\right)Y_{-\frac{3}{2}}(y)-g_1\left(\frac{\theta_*}{\theta}y\right)J_{-\frac{3}{2}}(y)\right|^2,
\end{split}
\label{eq_rhoBAFTER}
\end{equation}
where $g_1(x)$ and $g_2(x)$ are defined in App. \ref{app_usefulfunctions}. Finally, we have
\begin{equation}
\frac{\rho_{\text{B}}(\theta)}{M^4_{\text{Pl}}}=\frac{1}{M^4_{\text{Pl}}}\begin{cases}
\rho_{\text{B}}^{(\text{in})}(\theta) & \theta>\theta_* \\ \rho_{\text{B}}^{(\text{fin})}(\theta) & \theta<\theta_*
\end{cases}.
\label{eq_rhoBFULL}
\end{equation}
As a last contribution, we are left to compute $\rho_{\text{L}\perp}$. We find the following general expression for $\rho_{\text{L}\perp}$, valid for each $\alpha$ and independent of the choice of the initial conditions:
\begin{equation}
\rho_{\text{L}\perp}(\eta)=\frac{m^2}{2\pi^2a^2}\int_{k_i}^{k_H(\eta)}dk k^2\left|\frac{\mathcal{A}_{\lambda}}{f}\right|^2.
\label{eq_rhoLperpGENERAL}
\end{equation}
When we specialize Eq. (\ref{eq_rhoLperpGENERAL}) for the solutions in Eq. (\ref{eq_transvsolution}), valid before backreaction, we find
\begin{equation}
\begin{split}
\frac{\rho_{\text{L}\perp}^{(\text{in})}(\theta)}{M^4_{\text{Pl}}}&=\frac{1}{8\pi}\left(\frac{m}{M_{\text{Pl}}}\right)^2\left(\frac{H_I}{M_{\text{Pl}}}\right)^2\times \\&\times \left(\frac{\theta}{\theta_{\text{min}}}\right)^{2\alpha}\int_{\theta}^1dy y^2\left|H_{\alpha+\frac{1}{2}}^{(1)}(y)\right|^2.
\end{split}
\label{eq_rhoLperpINITIAL}
\end{equation}
Before of extending the result to the whole duration of inflation, it is important to discuss the massless limit of $\rho_{\text{L}\perp}$. This is not a gauge-invariant quantity, and thus by virtue of what discussed above the massless limit should be discontinuous. However, we can see that, in this case, sending $m\to 0$ yields $\rho_{\text{L}\perp}\to 0$. Even though this seems in contrast with our previous reasoning, in fact it is perfectly coherent. The key point is that $\rho_{\text{L}\perp}$ comes from transverse modes, that are physical both for $m=0$ and $m\neq 0$, while $\rho_0$ and $\rho_{\text{L}\parallel}$ are associated with longitudinal modes.
\begin{itemize}
\item[-] On the one hand, we have a contribution to the electromagnetic energy density, $\rho_{\text{L}\perp}$, that emerges only in the massive theory, but is related to transverse modes, that are physical also for $m=0$. Accordingly, there are no reasons for which, sending $m\to 0$, we should expect $\rho_{\text{L}\perp}\neq 0$, and indeed we get $\rho_{\text{L}\perp}\to 0$.
\item[-] On the other hand, we have two contributions to the electromagnetic energy density, $\rho_0$ and $\rho_{\text{L}\parallel}$, that emerge only in the massive theory too, but are related to longitudinal modes, that are unphysical for $m=0$. Accordingly, it is not sufficient to send $m\to 0$ to recover the massless case, but we also need to restore the gauge invariance of the theory. A confirmation of this claim is that, instead, a gauge-invariant quantity that depends on longitudinal modes, such as $\rho_{\text{E}\parallel}$, recovers the massless limit continuously.
\end{itemize}

\begin{figure}
\includegraphics[width=0.5\textwidth]{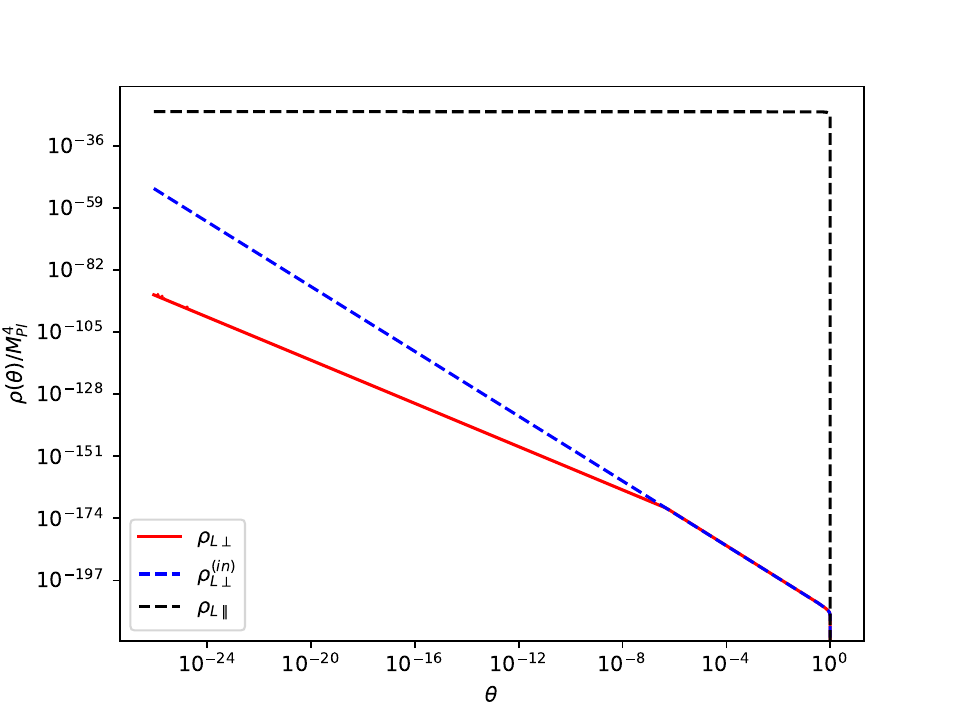}
\caption{Function $\rho_{\text{L}\perp}$ (red line) and $\rho_{\text{L}\perp}^{(\text{in})}$ (dashed blue line) in comparison, obtained with and without accounting for backreaction, respectively. The function $\rho_{\text{L}\parallel}$ (dashed black line) is plotted as well. In the plot, we consider $\frac{m}{M_{\text{Pl}}}=10^{-33}$, choosing the maximum value of the mass allowed within our framework, that is however much greater than the experimentally allowed range.}
\label{fig_rhoLperp}
\end{figure}

Then, we follow the usual prescription in order to extend the solution to the whole duration of the inflation. We name the solution valid after backreaction $\rho_{\text{L}\perp}^{(\text{fin})}$, and it turns out to be
\begin{equation}
\begin{split}
&\frac{\rho_{\text{L}\perp}^{(\text{fin})}(\theta)}{M^4_{\text{Pl}}}=\frac{1}{8\pi}\left(\frac{m}{M_{\text{Pl}}}\right)^2\left(\frac{H_I}{M_{\text{Pl}}}\right)^2\left(\frac{\theta_*}{\theta_{\text{min}}}\right)^{2\alpha}\left(\frac{\theta}{\theta_*}\right)^{-4}\times \\&\times \int_{\theta}^1dyy^2\left|d_1\left(\frac{\theta_*}{\theta}y\right)J_{-\frac{3}{2}}(y)+d_2\left(\frac{\theta_*}{\theta}y\right)Y_{-\frac{3}{2}}(y)\right|^2,
\end{split}
\label{eq_rhoLperpAFTER}
\end{equation}
where $d_1(x)$ and $d_2(x)$ are defined in App. \ref{app_usefulfunctions}. Finally, we have
\begin{equation}
\frac{\rho_{\text{L}\perp}(\theta)}{M^4_{\text{Pl}}}=\frac{1}{M^4_{\text{Pl}}}\begin{cases}
\rho_{\text{L}\perp}^{(\text{in})}(\theta) & \theta>\theta_* \\ \rho_{\text{L}\perp}^{(\text{fin})}(\theta) & \theta<\theta_*
\end{cases}.
\label{eq_rhoLperpFULL}
\end{equation}
We will prove that $\rho_{\text{L}\perp}$ is negligible with respect to its longitudinal counterpart, $\rho_{\text{L}\parallel}$, for each $\alpha<-2$. For this purpose, we show in Fig. \ref{fig_rhoLperp} the curve for $\alpha=-2.5$, with and without backreaction, further than its longitudinal counterpart, $\rho_{\text{L}\parallel}$. In particular, the plot shows that, for $\alpha=-2.5$, $\rho_{\text{L}\perp}\ll \rho_{\text{L}\parallel}$.

In addition, we can see that $\rho_{\text{L}\perp}$ grows as inflation proceeds, and in particular considering the backreaction regime the growth turns out to be more slowly. Straightforwardly, the maximum of the function $\rho_{\text{L}\perp}$ is located in $\theta_{\text{min}}$.

\begin{figure}
\includegraphics[width=0.5\textwidth]{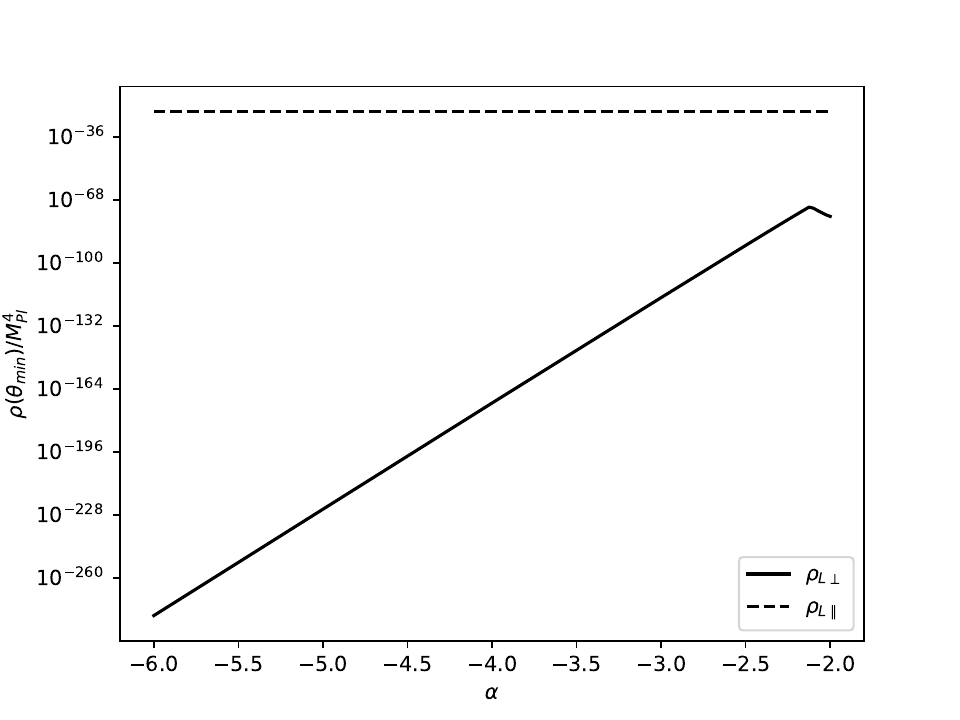}
\caption{Maximum values of $\rho_{\text{L}\perp}$ as a function of $\alpha$. For the sake of completeness, we also plot the plateau value of $\rho_{\text{L}\parallel}$, that is independent of $\alpha$.}
\label{fig_maxrhoLperpVSalpha}
\end{figure}

In order to verify that $\rho_{\text{L}\perp}$ remains lower than $\rho_{\text{L}\parallel}$ for each $\alpha$, it is sufficient to compute the maximum of $\rho_{\text{L}\perp}$, that is, $\rho_{\text{L}\perp}(\theta_{\text{min}})$, for different values of $\alpha$ and verify that each of them is lower than the plateau region of $\rho_{\text{L}\parallel}$. For this purpose, we should use
\begin{itemize}
\item[-] the functional form of $\rho_{\text{L}\perp}^{(\text{fin})}$ if $|\alpha|$ is sufficiently big to allow $\theta_*$ to exist, while
\item[-] $\rho_{\text{L}\perp}^{(\text{in})}$ otherwise.
\end{itemize}
Accordingly, we obtain the plot in Fig. \ref{fig_maxrhoLperpVSalpha}.

Fig. \ref{fig_maxrhoLperpVSalpha} shows that the maximum value of $\rho_{\text{L}\perp}$ grows when $\alpha$ goes from $-2$ until $\simeq -2.15$. Indeed, in this range of values, Eq. (\ref{eq_thetastarVSalpha}) cannot be satisfied for any $\theta_*\in[\theta_{\text{min}},1]$, and therefore backreaction does not occur. Accordingly, having a bigger modulus of $\alpha$ means having a faster growth of $\rho_{\text{L}\perp}$, that will reach a higher maximum. Differently, when $\alpha$ further grows (in modulus), we observe a net reduction of the maximum value reached by $\rho_{\text{L}\perp}$. This happens because, even though the initial growth of $\rho_{\text{L}\perp}$ is faster, $\theta_*$ is closer to $1$ and backreaction occurs earlier. Accordingly, the curve becomes flatten earlier and reaches a lower maximum.

In any case, for each $\alpha$ the maximum of $\rho_{\text{L}\perp}$ remains various orders of magnitude lower than the maximum of $\rho_{\text{L}\parallel}$. Accordingly, we can safely consider $\rho_{\text{L}\parallel}\gg \rho_{\text{L}\perp}$, that is, $\rho_{\text{L}}=\rho_{\text{L}\parallel}+\rho_{\text{L}\perp}\simeq \rho_{\text{L}\parallel}$.

\begin{figure}
\includegraphics[width=0.5\textwidth]{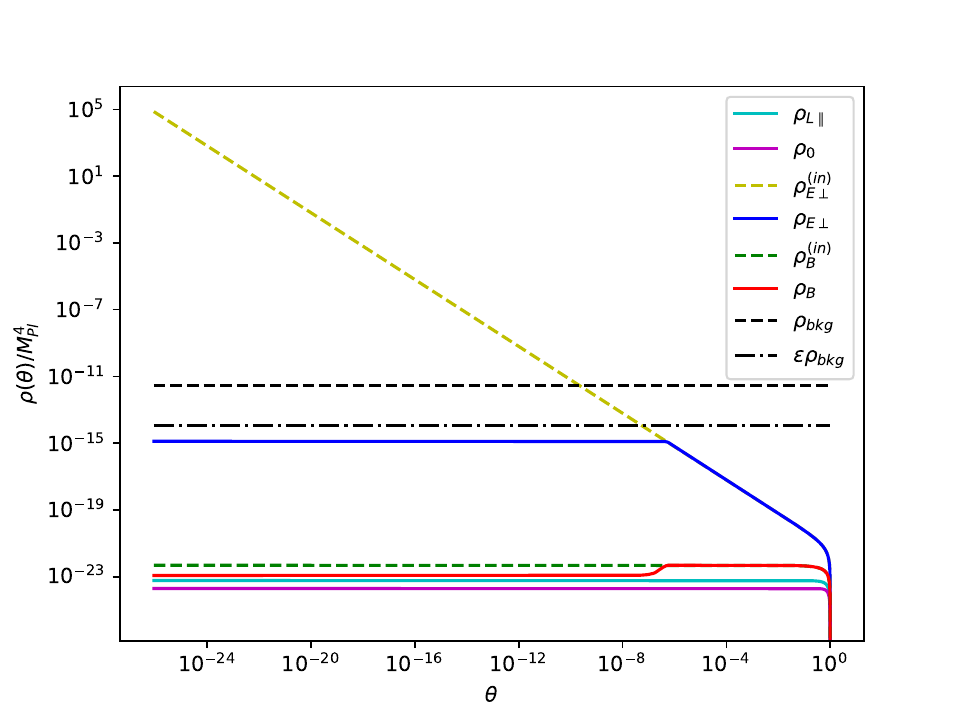}
\caption{Main contributions to the electromagnetic energy density. The electric and magnetic components are plotted with and without accounting for the backreaction. In particular, the backreaction is considered in $\rho_{\text{E}\perp}$ (blue line) and $\rho_{\text{B}}$ (red line), while it is not considered in $\rho_{\text{E}\perp}^{(\text{in})}$ (yellow dashed line) and $\rho_{\text{B}}^{(\text{in})}$ (dashed green line). The longitudinal components, $\rho_0$ (purple line) and $\rho_{\text{L}\parallel}$ (light-blue line), are plotted as well.}
\label{fig_rhoEMcontributions}
\end{figure}

Finally, in Fig. \ref{fig_rhoEMcontributions} we plot the relevant contributions to the electromagnetic energy density: $\rho_{\text{E}\perp}$, $\rho_{\text{B}}$, $\rho_0$ and $\rho_{\text{L}\parallel}$. In particular, the first two are plotted with and without accounting for backreaction, while the other do not change among the two cases.

As the small mass of the vector fields does not affect significantly the transverse mode dynamics, the electric and magnetic components of the energy density remain almost the same as obtained for $m=0$. Accordingly, the electric contribution is dominant, while the magnetic component behaves as a perturbation.

Then, we consider the contributions coming from the longitudinal modes. These emerge as new energy components with respect to the massless scenario, and it is thus interesting to understand their nature. In particular, the longitudinal contributions turn out to be comparable to the magnetic component, providing only small corrections to the electric background. This is a consequence of the small mass of the vector field, that is highly constrained by the observations.

The enormous gap in terms of magnitude between the electric energy density and the other components enables us to consider the electric energy density as the background, while the other components as perturbations. As long as this separation is kept, we can deal with the electric background and the perturbative contributions independently. Specifically, we will follow this approach in the Sect. \ref{Schwinger}, where we will introduce the Schwinger effect.

\subsection{The magnetic field at our time}

Before of introducing the Schwinger effect in the model, it is interesting to compute the value of the magnetic field at our time. For this purpose, we need the value of the magnetic energy density at the end of inflation. According to the red curve in Fig. \ref{fig_rhoEMcontributions}, we get
\begin{equation}
\frac{\rho_{\text{B}}(\theta_{\text{min}})}{M^4_{\text{Pl}}}\simeq 1.18\times 10^{-23}.
\label{eq_rhoB-end-inflation}
\end{equation}
From it, we can compute the magnetic field at the end of the inflation,
\begin{equation}
B_e\equiv B(\theta_{\text{min}})=\sqrt{\frac{2\rho_B(\theta_{\text{min}})}{f^2(\theta_{\text{min}})}}\simeq 4.86\times 10^{-12}M^2_{\text{Pl}},
\label{eq_B-end-inflation}
\end{equation}
where we use $f(\theta_{\text{min}})=1$. Finally, the magnetic field at our time, say $B_0$, can be computed from $B_e$ through the magnetic flux conservation law, implying that $B\propto a^{-2}$. Accordingly, we get
\begin{equation}
B_0\equiv \left(\frac{a_e}{a_0}\right)^2B_e\simeq 2.67\times 10^{-17}\text{ G},
\label{eq_B-today}
\end{equation}
where we conventionally use $a_0=1$, $a_e=-\frac{1}{H_I\eta_f}$, and consider that $1\text{ G}\simeq 1.95\times 10^{-20}\text{ GeV}^2$.

Comparing our result with the experimental bounds discussed in Sect. \ref{intro}, namely, $10^{-18}\le B_0\le 10^{-9}\text{ G}$, we point out that our model predicts a magnetic field value perfectly compatible with observational data. Moreover, we remark that this value is not significantly affected by the presence of a nonzero mass of the vector field, as long as the latter is small compatibly with experimental bounds.

\section{The Schwinger effect}{\label{Schwinger}}

By virtue of the outcomes obtained in Sect. \ref{energy_density}, one may think that, nowadays, the electric fields are observed in larger amounts with respect to magnetic ones, as the former turned out to be much more present than the latter at the end of inflation.

However, experimental observations do not describe this scenario at all. Conversely, nowadays, only the magnetic fields are detected, while no electric fields are measured throughout the universe. In order to explain the observations, we need to find a mechanism to suppress the strong electric fields produced during inflation.

For this purpose, it is commonly invoked the Schwinger effect, consisting in the production of particle pairs by vacuum in presence of strong electric fields. In inflationary magnetogenesis, strong electric fields are present during the final stages of inflation, and particles should be thus produced via the Schwinger mechanism. Therefore, including the Schwinger effect in the model, further than to explain the observations, is necessary in order to formulate a more complete and faithful model.

In this work, we consider the Schwinger production of massless charged scalar particles. In this regard, we consider a quantum complex scalar field $\chi$, with Lagrangian density
\begin{equation}
\begin{split}
\mathcal{L}_{\text{Schw}}&=\sqrt{-g}\left(D_{\mu}\chi\right)\left(D^{\mu}\chi\right)^*=\sqrt{-g}\Big[\left(\partial_{\mu}\chi\right)\left(\partial^{\mu}\chi^*\right)+\\&+e^2A^{\mu}A_{\mu}\chi\chi^*+ieA_{\mu}\left(\chi\partial^{\mu}\chi^*-\chi^*\partial^{\mu}\chi\right)\Big].
\end{split}
\label{eq_SchwingerLagrangian}
\end{equation}
Eq. (\ref{eq_SchwingerLagrangian}) is \emph{de facto} the theory of the scalar QED, where the complex scalars are coupled to the electromagnetic field through the minimal prescription. Accordingly, we study the following Lagrangian model:
\begin{equation}
\mathcal{L}=\mathcal{L}_{\text{m}}+\mathcal{L}_{\text{Schw}}.
\label{eq_LagrangianFULL}
\end{equation}
Defining the current\footnote{We will see that this is not the whole conserved current in the scalar sector. The latter is in fact the quantity in round brackets in Eq. (\ref{eq_conservationlawscalar}).}
\begin{equation}
J_{\mu}\equiv i\left(\chi\partial_{\mu}\chi^*-\chi^*\partial_{\mu}\chi\right),
\label{eq_current}
\end{equation}
the electromagnetic equations turn out to be
\begin{equation}
\partial_{\mu}F^{\mu\nu}+\left(3H+2\frac{\dot{f}}{f}\right)F^{0\nu}+\left(\frac{m^2+2e^2\chi\chi^*}{f^2}\right)A^{\nu}=-\frac{e}{f^2}J^{\mu}.
\label{eq_EMequationSCHWINGER}
\end{equation}
On the other hand, the equation for the scalar field $\chi$ reads
\begin{equation}
\ddot{\chi}-\frac{\nabla^2\chi}{a^2}+3H\dot{\chi}+2ieA^{\mu}\partial_{\mu}\chi-e^2A^{\mu}A_{\mu}\chi=0.
\label{eq_EOMchi}
\end{equation}
We immediately notice that the fully quantum approach followed for $e=0$ cannot be pursued here, as we should exactly solve an interacting quantum theory. However, the results obtained for $e=0$ suggest that we can deal with the background electric energy density and the other perturbative components independently. This point is fundamental, and constitutes the bridge between the frameworks with $e=0$ and $e\neq 0$.
\begin{itemize}
\item[-] The Schwinger particle production is a nonperturbative effect, thus it only affects the background dynamics. In our framework, this is governed by the transverse electric fields, as long as the electric energy density remains dominant with respect to the other components.
\item[-] The transverse electric fields are associated with the time variation of the electromagnetic field. Accordingly, they can be described consistently by classical homogeneous electromagnetic fields. In particular, we expect that, for $e=0$, considering a classical homogeneous electromagnetic field yields the same curve for the electric energy density as plotted in Fig. \ref{fig_rhoEMcontributions}. We will check this explicitly in Sect. \ref{subverifye=0}.
\item[-] Therefore, for non-null couplings, we can describe the background dynamics considering classical homogeneous electromagnetic fields and quantum complex scalar fields. Remarkably, this semiclassical approach is consistent as long as the electromagnetic background is dominated by the transverse electric fields.
\item[-] Eq. (\ref{eq_EOMchi}) will be solved in presence of a homogeneous classical electromagnetic field. This is a very important point, as if both the fields -- scalar and electromagnetic -- were quantized, the resolution would be considerably more involved.
\item[-] On the other hand, Eq. (\ref{eq_EMequationSCHWINGER}) will be solved for a homogeneous classical electromagnetic field. There, in lieu of $\chi\chi^*$ and $J^\mu$, we should take the respective VEVs, and these describe the backreaction of the produced scalar particles on the electric energy density.
\item[-] All the perturbative components of the electromagnetic energy density, that is, longitudinal and magnetic, remain unaffected by the Schwinger particle production. Accordingly, they continue to be described by quantum fields, and their functional forms remain the same as plotted in Fig. \ref{fig_rhoEMcontributions}, obtained for $e=0$.
\end{itemize}

\subsection{Verifying the consistency of the semiclassical approach with the limit $e=0$}{\label{subverifye=0}}

For the sake of consistency, we need to check if, for $e=0$, a homogeneous classical electromagnetic field can really reproduce the electric energy density plotted in Fig. \ref{fig_rhoEMcontributions}, obtained within the fully quantum approach.

When we consider a homogeneous classical electromagnetic field, no distinction between longitudinal and transverse modes makes sense anymore. Therefore, we consider Eq. (\ref{eq_EMequation}) and specialize it for $\nu=0$ and $\nu=i$, obtaining
\begin{subequations}
\begin{align}
&A_0=0,
\label{eq_GaussLawCLASSICAL}\\
&\ddot{{\bf A}}+\left(H+2\frac{\dot{f}}{f}\right)\dot{\bf A}+\frac{m^2}{f^2}{\bf A}=0,
\label{eq_spatialEMCLASSICAL}
\end{align}
\end{subequations}
respectively. We notice that Eq. (\ref{eq_spatialEMCLASSICAL}) resembles Eq. (\ref{eq_EMtransverse}), without the Laplacian term. Therefore, we can deduce that only transverse modes are described within the classical framework. Accordingly, describing the transverse electric fields through classical electromagnetic fields turns out to be perfectly consistent.

In conformal time, defining $\mathcal{A}_i=fA_i$, Eq. (\ref{eq_spatialEMCLASSICAL}) becomes
\begin{equation}
\mathcal{A}_i''+\left(\frac{a^2m^2}{f^2}-\frac{\alpha(\alpha+1)}{\eta^2}\right)\mathcal{A}_i=0.
\label{eq_spatialEMCLASSICALconformal}
\end{equation}
Here too, we can neglect the positive mass term with respect to the tachyonic negative one. Indeed, $\frac{a^2m^2}{f^2}<a^2m^2$, as $f>1$. Therefore, in order to establish which term is dominant, we should compare $\frac{m^2}{H_I^2}$ and $\alpha(\alpha+1)$: the latter is of the order of the unity, while the former is enormously smaller, and can be thus neglected. Therefore, we get
\begin{equation}
\mathcal{A}_i''-\frac{\alpha(\alpha+1)}{\eta^2}\mathcal{A}_i=0.
\label{eq_spatialEMCLASSICALconformalNOmass}
\end{equation}
Solving for $\mathcal{A}_i$, the original $A_i$ turns out to be
\begin{equation}
A_i(\eta)=\frac{1}{f}\left[c_1(-\eta)^{\alpha+1}+c_2(-\eta)^{-\alpha}\right],
\label{eq_CLASSICALsolutionEM}
\end{equation}
with $c_1$ and $c_2$ free constants to be determined. Accordingly, the electric field is
\begin{equation}
\left({\bf E}\right)_i=\frac{\dot{A_i}}{a}=\frac{A_i'}{a^2}=-\frac{c_1(2\alpha+1)}{a^2f}(-\eta)^{\alpha},
\label{eq_CLASSICALelectricfield}
\end{equation}
and for the electric energy density we obtain
\begin{equation}
\frac{\rho_{\text{E}}^{(\text{in})}(\theta)}{M^4_{\text{Pl}}}=\frac{|{\bf d_1}|^2(2\alpha+1)^2}{2}\left(\frac{H_I}{M_{\text{Pl}}}\right)^4\theta^{2\alpha+4},
\label{eq_CLASSICALrhoE}
\end{equation}
with ${\bf d_1}\equiv\frac{{\bf c_1}}{k_{\text{i}}^{\alpha+2}}$.

We can check that it is not possible to set $\rho^{(\text{in})}_{\text{E}}(\theta=1)=0$, as we had in the fully quantum scenario. However, there, the integral over the modes was constructed such to naturally give $\rho^{(\text{in})}_{\text{E}}(\theta=1)=0$, as a consequence of considering only super-Hubble modes. That is, it was a consequence of the cutoff choice.

Overall, we are more interested in the behavior during the inflation, that instead should be reproduced accurately. Accordingly, we conveniently choose the boundary conditions such that the two scenarios coincide in $\theta_*$, yielding $|{\bf d_1}(\alpha)|^2=\frac{4.53\times 10^7}{4\pi(2\alpha+1)^2}\theta_*^{-2\alpha-4}(\alpha)$.

\begin{figure}
\includegraphics[width=0.5\textwidth]{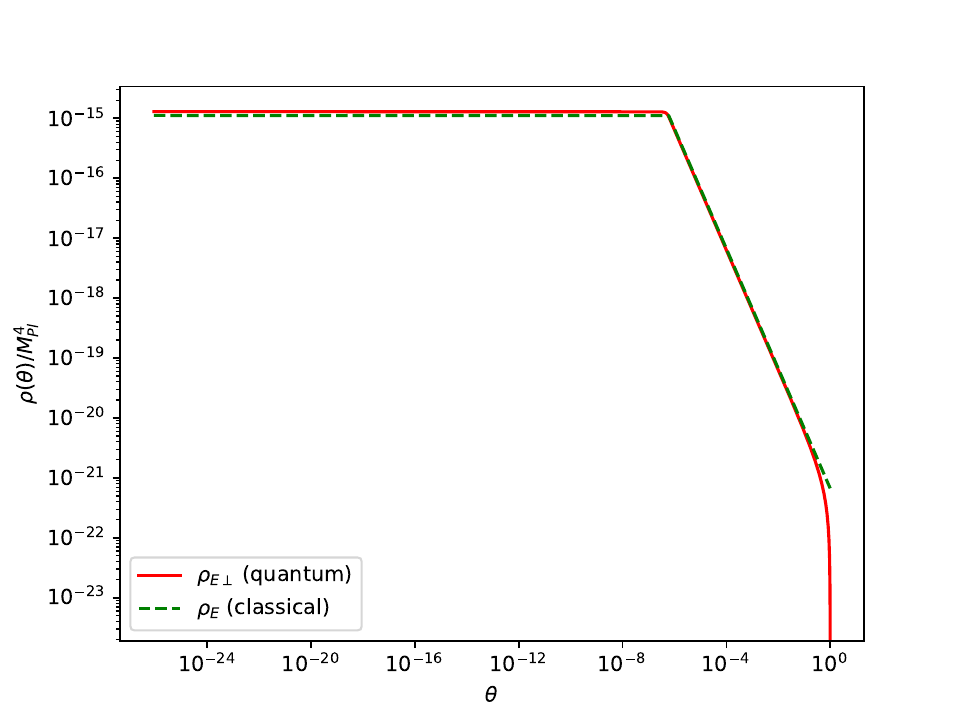}
\caption{Eqs. (\ref{eq_rhoECLASSICALfull}) and (\ref{eq_rhoEperpFULL}) in comparison, obtained through a semiclassical (dashed green line) and a quantum (red line) approach, respectively.}
\label{fig_quantumVSclassical}
\end{figure}

Then, we need to address the backreaction problem. However, in this framework this is rather simple, as it is enough to keep the solution constant for $\theta>\theta_*$. Accordingly, from a classical perspective, we obtain the following expression for the electric energy density:
\begin{equation}
\frac{\rho_{\text{E}}(\theta)}{M^4_{\text{Pl}}}=\left(\frac{H_I}{M_{\text{Pl}}}\right)^4\frac{|{\bf d_1}(\alpha)|^2(2\alpha+1)^2}{2}\begin{cases}
\theta^{2\alpha+4} & \theta>\theta_*\\ \theta_*^{2\alpha+4} & \theta<\theta_*
\end{cases}.
\label{eq_rhoECLASSICALfull}
\end{equation}
To check that our semiclassical approach is consistent with the fully quantum scenario, we need to check that the electric energy densities obtained in the two ways are the same. In particular, we should compare Eqs. (\ref{eq_rhoECLASSICALfull}) and (\ref{eq_rhoEperpFULL}). The two curves are plotted in Fig. \ref{fig_quantumVSclassical}.

As discussed previously, close to $\theta=1$ the two curves differ. However, we are mainly interested in the behavior during inflation. There, we notice an almost perfect agreement between the two curves, apart from a small discrepancy during the backreaction regime. Remarkably, this has nothing to do with physics, but rather it is something related to our prescription for dealing with the backreaction problem; a numerical approach to the problem would simply solve the question.

Indeed, this small discrepancy is due to the fact that Eq. (\ref{eq_rhoECLASSICALfull}) is constructed such to guarantee continuity between the two regimes in $\theta_*$, but not differentiability. This is perfectly expected, as long as only one free parameter appears within Eq. (\ref{eq_CLASSICALrhoE}), and thus only one of the two conditions can be imposed. As it is not differentiable in $\theta_*$, the curve obtained classically flattens abruptly, reaching the plateau earlier and thus taking a lower asymptotic value.

Overall, the semiclassical approach works well in reproducing the electric energy density for $e=0$. Accordingly, we can introduce the Schwinger effect within a classical electric background, and study how the latter is affected by the particle production. We remark that the longitudinal and magnetic components remain unaffected by the Schwinger particle production, and their curves remain those in Fig. \ref{fig_rhoEMcontributions}.

\subsection{Schwinger particle production and backreaction on the electric background}

In order to study how the electric background is affected by the Schwinger particle production, we consider Eq. (\ref{eq_EMequationSCHWINGER}) for a homogeneous classical electromagnetic field, and replace $\chi\chi^*$ and $J^\mu$ with their VEVs. In particular, for $\nu=0$ and $\nu=i$ we get
\begin{subequations}
\begin{align}
&\left(m^2+2e^2\braket{\chi\chi^*}\right)A_0=-e\braket{J_0},
\label{eq_GaussSchwingerCLASSICAL}\\
&\ddot{A_i}+\left(H+2\frac{\dot{f}}{f}\right)\dot{A_i}+\left(\frac{m^2+2e^2\braket{\chi\chi^*}}{f^2}\right)A_i=-\frac{e}{f^2}\braket{J_i},
\label{eq_spatialSchwingerCLASSICAL}
\end{align}
\end{subequations}
respectively. We can check that for $e=0$ we recover Eqs. (\ref{eq_GaussLawCLASSICAL}) and (\ref{eq_spatialEMCLASSICAL}). In particular, we notice that the Gauss law for $e=0$ yields $A_0=0$, while for $e\neq0$ at first glance this does not seem guaranteed.

However, this condition can be recovered when $e\neq 0$ too. Indeed, combining the divergence of the electromagnetic equations of motion,
\begin{equation}
e\nabla_{\mu}\left(J^{\mu}+2e\chi\chi^*A^{\mu}\right)+m^2\nabla_{\mu}A^{\mu}=0,
\label{eq_divEOMelectromagnetic}
\end{equation}
with the conservation law for the scalar sector,
\begin{equation}
\nabla_{\mu}\left(J^{\mu}+2e\chi\chi^*A^{\mu}\right)=0,
\label{eq_conservationlawscalar}
\end{equation}
we straightforwardly obtain
\begin{equation}
\nabla_{\mu}A^{\mu}=0\implies A_0(t)=A_0(0)e^{-3Ht}.
\label{eq_A0Eq}
\end{equation}
Setting $A_0(0)=0$, we naturally recover the identity $A_0=0$. From a physical viewpoint, the choice of such an initial condition is driven mainly by two reasons.
\begin{itemize}
\item[-] The initial value of $A_0$ should not depend on the fact that we consider a non-null coupling or a null one. Indeed, the interaction with the scalar field should start to be relevant during the final stages of inflation, when the charged scalar particles have been produced in a significant amount. For $e=0$, $A_0=0$ identically, and therefore also $A_0(0)=0$. The same initial condition should be thus set also for $e\neq 0$.
\item[-] Unless we set $A_0(0)=0$, $A_0(t)$ turns out to be a decreasing function of the cosmic time, that is not consistent with the result found within the fully quantum approach.
\end{itemize}
Accordingly, we proceed setting $A_0(0)=0$ and therefore we get $A_0=0$ identically also for $e\neq 0$. Furthermore, we notice that the value of $A_0$ does not alter the dynamics of $A_i$ when the electromagnetic field is dealt with as a classical field, and therefore our choice does not modify the background dynamics and the future results.

About the quantum complex scalar field, we conventionally write it through its mode expansion,
\begin{equation}
\hat{\chi}(t,{\bf x})=\frac{1}{a}\int\frac{d^3k}{(2\pi)^{\frac{3}{2}}}\left[\hat{c}_{{\bf k}}\xi_{\bf k}(t)e^{i{\bf k}\cdot{\bf x}}+\hat{d}^{\dag}_{{\bf k}}\xi_{\bf k}^*(t)e^{-i{\bf k}\cdot{\bf x}}\right],
\label{eq_quantumchi}
\end{equation}
with $\left[\hat{c}_{\bf k},\hat{c}^{\dag}_{\bf p}\right]=\left[\hat{d}_{\bf k},\hat{d}^{\dag}_{\bf p}\right]=\delta^{(3)}\left({\bf k}-{\bf p}\right)$. 

It is convenient to rewrite Eq. (\ref{eq_spatialSchwingerCLASSICAL}) in conformal time in terms of $\mathcal{A}_i=fA_i$, yielding
\begin{equation}
\mathcal{A}_i''+\left[\frac{a^2m^2}{f^2}+2\frac{e^2}{f^2}\braket{\xi\xi^*}-\frac{\alpha(\alpha+1)}{\eta^2}\right]\mathcal{A}_i=-\frac{e}{f}\braket{\mathcal{J}_i},
\label{eq_spatialSchwingerCLASSICALconformal}
\end{equation}
where $\braket{\xi\xi^*}\equiv a^2\braket{\chi\chi^*}$ and $\braket{\mathcal{J}_i}=a^2\braket{J_i}$.

On the other hand, from Eq. (\ref{eq_EOMchi}), we obtain the following mode equations with respect to conformal time:
\begin{equation}
\xi_{\bf k}''+\left[\left(k_i+\frac{e}{f}\mathcal{A}_i\right)^2-\frac{2}{\eta^2}\right]\xi_{\bf k}=0.
\label{eq_modeequations}
\end{equation}
As done for passing from Eq. (\ref{eq_spatialEMCLASSICALconformal}) to Eq. (\ref{eq_spatialEMCLASSICALconformalNOmass}), we neglect the mass term with respect to the tachyonic negative one in Eq. (\ref{eq_spatialSchwingerCLASSICALconformal}). In addition, in this case we have a dynamical mass correction, proportional to $\braket{\xi\xi^*}$. In fact, this is not only a correction, rather it becomes dominant as the inflation proceeds and scalar particles are produced via the Schwinger effect. Moreover, we notice that an effective coupling is naturally defined,
\begin{equation}
e_{\text{eff}}(\eta)\equiv \frac{e}{f(\eta)},
\label{eq_effectivecoupling}
\end{equation}
and it grows as the inflation proceeds. This fact strengthens the increasing relevance of the dynamical mass correction due to the Schwinger particle production during the inflation, meaning that the Schwinger effect becomes more relevant during the final stages of inflation.

We can arbitrarily consider $\mathcal{A}_i\equiv\left(0,0,\mathcal{A}_i\right)$, and accordingly we finally write the following system of coupled integro-differential equations:
\begin{equation}
\begin{cases}
\xi_{\bf k}''+\left[k^2+2k_ze_{\text{eff}}\mathcal{A}_i+e^2_{\text{eff}}\mathcal{A}_i^2-\frac{2}{\eta^2}\right]\xi_{\bf k}=0 \\ \mathcal{A}_i''+\left[2e^2_{\text{eff}}\braket{\xi\xi^*}-\frac{\alpha(\alpha+1)}{\eta^2}\right]\mathcal{A}_i=-e_{\text{eff}}\braket{\mathcal{J}_i}
\end{cases}.
\label{eq_systemSchwinger}
\end{equation}
The VEVs in question are
\begin{subequations}
\begin{align}
&\braket{\xi\xi^*}=\int\frac{d^3k}{(2\pi)^3}\xi_{\bf k}\xi^*_{\bf k}=\frac{1}{4\pi^2}\int_{-1}^1dx\int_{k_{\text{i}}}^{k_H(\eta)}k^2|\xi_{\bf k}|^2dk,
\label{eq_VEVmass}\\
&\braket{\mathcal{J}_i}=2\int\frac{d^3k}{(2\pi)^3}k_z\xi_{\bf k}\xi^*_{\bf k}=\frac{1}{2\pi^2}\int_{-1}^1dx\int_{k_{\text{i}}}^{k_H(\eta)}k^3x|\xi_{\bf k}|^2dk.
\label{eq_VEVcurrent}
\end{align}
\end{subequations}
We stress that the momentum range entering the above relations follows the same effective prescription as before. Accordingly, the upper limit $k_H(\eta)$ is not completely consistent as a UV cutoff for the entire scalar quantum theory, simply because, within the present semiclassical description, we retain the long-wavelength charged sector that contributes to the backreaction on the effective electric background. Hence, $\langle\xi\xi^*\rangle$ and $\langle \mathcal{J}_i\rangle$, appearing in the following equations, are coarse-grained expectation values restricted to this sector. Consequently, complete renormalized expectation values, obtained by integrating over the full momentum spectrum, can be obtained in the context of quantum gravity scenarios. 

Finally, Eq. (\ref{eq_systemSchwinger}) turns to
\begin{equation}
\begin{cases}
\xi_{\bf k}''+\left[k^2+2kxe_{\text{eff}}\mathcal{A}_i+e^2_{\text{eff}}\mathcal{A}_i^2-\frac{2}{\eta^2}\right]\xi_{\bf k}=0 \\ \begin{split}
\mathcal{A}_i''+&\left[\frac{e^2_{\text{eff}}}{2\pi^2}\int_{-1}^1dx\int_{k_{\text{i}}}^{k_H}k^2|\xi_{\bf k}|^2dk-\frac{\alpha(\alpha+1)}{\eta^2}\right]\mathcal{A}_i=\\&=-\frac{e_{\text{eff}}}{2\pi^2}\int_{-1}^1dx\int_{k_{\text{i}}}^{k_H}k^3x|\xi_{\bf k}|^2dk
\end{split}
\end{cases}\!\!\!\!.
\label{eq_SchwingerbkgDYNAMICS}
\end{equation}

\section{Numerical resolution of the coupled dynamics}

The system in Eq. (\ref{eq_SchwingerbkgDYNAMICS}) should be solved numerically. Accordingly, it is useful to rewrite everything in terms of dimensionless quantities,
\begin{equation}
\tilde{\mathcal{A}}_i\equiv\frac{\mathcal{A}_i}{k_{\text{i}}}\quad \tilde{\xi}_{\bf k}\equiv \sqrt{k_{\text{i}}}\xi_{\bf k},\quad \tilde{k}\equiv\frac{k}{k_{\text{i}}},
\label{eq_dimensionlessquantities}
\end{equation}
and with respect to the dimensionless variable $\theta$. We get
\begin{equation}
\begin{cases}
\frac{d^2\tilde{\xi}_{\bf k}}{d\theta^2}+\left[\tilde{k}^2+2\tilde{k}xe_{\text{eff}}\tilde{\mathcal{A}_i}+e^2_{\text{eff}}\tilde{\mathcal{A}}_i^2-\frac{2}{\theta^2}\right]\tilde{\xi}_{\bf k}=0 \\ \begin{split}
\frac{d^2\tilde{\mathcal{A}}_i}{d\theta^2}+&\left[\frac{e^2_{\text{eff}}}{2\pi^2}\int_{-1}^1dx\int_1^{\tilde{k}_H}\tilde{k}^2|\tilde{\xi}_{\bf k}|^2d\tilde{k}-\frac{\alpha(\alpha+1)}{\theta^2}\right]\tilde{\mathcal{A}}_i=\\&=-\frac{e_{\text{eff}}}{2\pi^2}\int_{-1}^1dx\int_1^{\tilde{k}_H}\tilde{k}^3x|\tilde{\xi}_{\bf k}|^2d\tilde{k}
\end{split}
\end{cases}\!\!\!\!,
\label{eq_SchwingersystemDIMENSIONLESS}
\end{equation}
with $\tilde{k}_H(\theta)=\frac{1}{\theta}$. However, the system is still complicated to be addressed numerically, as the independent variable $\theta$ spans almost $27$ orders of magnitude. An efficient numerical simulation requires a more compact grid, that can be easily obtained by performing the following coordinate transformations,
\begin{equation}
N\equiv -\ln(\theta),
\label{eq_efoldinsnumber}
\end{equation}
considering \emph{de facto} the e-foldings number $N$ as the independent variable. In so doing, our independent variable lies within the interval $[0,-\ln(\theta_{\text{min}})]\simeq [0,60]$, and the grid is thus much more compact. In addition, it is also useful to define the new integration variable
\begin{equation}
v\equiv \theta \tilde{k}=e^{-N}\tilde{k}.
\label{eq_vintegrvariable}
\end{equation}
Accordingly, defining the functions
\begin{equation}
\begin{split}
&\delta(N)\equiv e^{-N}e_{\text{eff}}(N),\quad \sigma(N)\equiv \frac{e^{\frac{3N}{2}}\delta(N)}{\pi},\quad \\&\zeta(N)\equiv \frac{e^{3N}\delta(N)}{\pi^2}=\frac{\sigma^2(N)}{\delta(N)},
\end{split}
\label{eq_numericaleffectivecoupling}
\end{equation}
we obtain
\begin{equation}
\begin{cases}
\frac{d^2\tilde{\xi}_{\bf k}}{dN^2}+\frac{d\tilde{\xi}_{\bf k}}{dN}+\left[v^2+2vx\delta\tilde{\mathcal{A}_i}+\delta^2\tilde{\mathcal{A}}_i^2-2\right]\tilde{\xi}_{\bf k}=0 \\ \begin{split}
\frac{d^2\tilde{\mathcal{A}}_i}{dN^2}\!&+\!\frac{d\tilde{\mathcal{A}}_i}{dN}\!+\!\left[\frac{\sigma^2}{2}\int_{-1}^1\!\!\!\!dx\!\int_{e^{-N}}^1\!\!\!\!\!v^2|\tilde{\xi}_{\bf k}|^2dv-\alpha(\alpha+1)\right]\tilde{\mathcal{A}}_i=\\&=-\frac{\zeta}{2}\int_{-1}^1dx\int_{e^{-N}}^1v^3x|\tilde{\xi}_{\bf k}|^2dv
\end{split}
\end{cases}\!\!\!\!.
\label{eq_numericalSCHWINGERsystemV}
\end{equation}
We still notice that the numerical simulation could be hard, as the integration variable $v$ lies within the dynamical interval $[e^{-N},1]$, that can even span $\sim 27$ orders of magnitude. Accordingly, we define
\begin{equation}
y\equiv -\ln(v),
\label{eq_dimensionlessintegrvariablefinal}
\end{equation}
and Eq. (\ref{eq_numericalSCHWINGERsystemV}) turns to
\begin{equation}
\begin{cases}
\frac{d^2\tilde{\xi}_{\bf k}}{dN^2}+\frac{d\tilde{\xi}_{\bf k}}{dN}+\left[e^{-2y}+2e^{-y}x\delta\tilde{\mathcal{A}_i}+\delta^2\tilde{\mathcal{A}}_i^2-2\right]\tilde{\xi}_{\bf k}=0 \\ \begin{split}
\frac{d^2\tilde{\mathcal{A}}_i}{dN^2}\!&+\!\frac{d\tilde{\mathcal{A}}_i}{dN}\!+\!\left[\frac{\sigma^2}{2}\!\int_{-1}^1\!\!\!\!dx\!\int_0^N\!\!\!\!e^{-3y}|\tilde{\xi}_{\bf k}|^2dy-\alpha(\alpha+1)\right]\!\tilde{\mathcal{A}}_i=\\&=-\frac{\zeta}{2}\int_{-1}^1dx\int_{0}^Ne^{-4y}x|\tilde{\xi}_{\bf k}|^2dy
\end{split}
\end{cases}\!\!\!\!\!.
\label{eq_numericalSCHWINGERsystemY}
\end{equation}
In this way, both the independent variable $N$ and the integration variable $y$ span no more than $2$ orders of magnitude. From a numerical viewpoint, however, the system is still rather unstable. In particular, the integral on the right-hand side of the electromagnetic equation needs a very small integration step in order to provide stable results, showing otherwise frequent unphysical sign flips. This behavior is due to the fact that the scalar modes $\tilde{\xi}_{\bf k}$ are almost constant in $x$. Accordingly, the integrand on the right-hand side of the electromagnetic equation is almost odd in $x$, and when integrating it in the symmetric domain $[-1,1]$ at leading order we should obtain exactly zero, while a first-order correction should provide the biggest contribution to the integral. When the integration step is not small enough in order to correctly detect the cancellation of the leading-order contributions, a numerical instability appears.

\subsection{Numerical optimization of the system through perturbation theory}

We address this problem writing the scalar modes $\tilde{\xi}_{\bf k}$ as the sum of a leading-order contribution, $\tilde{\xi}_{\bf k}^{(0)}$, independent of $x$, and a small first-order correction, $\delta\tilde{\xi}_{\bf k}$, containing all the dependence on $x$:
\begin{equation}
\tilde{\xi}_{\bf k}=\tilde{\xi}_{\bf k}^{(0)}+\delta\tilde{\xi}_{\bf k}.
\label{eq_scalarmodesperturbation}
\end{equation}
Accordingly, the equation for the scalar modes splits into two different equations:
\begin{subequations}
\begin{align}
&\frac{d^2\tilde{\xi}^{(0)}_{\bf k}}{dN^2}+\frac{d\tilde{\xi}_{\bf k}^{(0)}}{dN}+\left[e^{-2y}+\delta^2\tilde{\mathcal{A}}_i^2-2\right]\tilde{\xi}_{\bf k}^{(0)}=0,
\label{eq_csi0}\\
&\frac{d^2\delta\tilde{\xi}_{\bf k}}{dN^2}+\frac{d\delta\tilde{\xi}_{\bf k}}{dN}+\left[e^{-2y}+\delta^2\tilde{\mathcal{A}}_i^2-2\right]\delta\tilde{\xi}_{\bf k}=-2e^{-y}x\delta\tilde{\mathcal{A}_i}\tilde{\xi}_{\bf k}^{(0)}.
\label{eq_deltacsi}
\end{align}
\end{subequations}
This perturbative approach is useful in order to write the integral on the right-hand side of the electromagnetic equation such that it is numerically stable. Indeed, in this way we can account for the cancellation of the leading-order contribution analytically, eliminating the instability problem. Indeed, the biggest contribution to this integral is given by the first-order correction, yielding
\begin{equation}
\begin{split}
\int_{-1}^1dx\int_0^N&e^{-4y}x|\tilde{\xi}_{\bf k}|^2dy\to\\& \to 2\int_{-1}^1dx\int_0^Ne^{-4y}x\text{Re}\left\{\tilde{\xi}_{\bf k}^{(0)*}\delta\tilde{\xi}_{\bf k}\right\}dy.
\end{split}
\label{eq_VEVcurrentperturbative}
\end{equation}
Conversely, the integral on the left-hand side of the electromagnetic equation has a nonzero leading-order contribution, and can be thus written simply as
\begin{equation}
\int_{-1}^1dx\int_0^Ne^{-3y}|\tilde{\xi}_{\bf k}|^2dy\to\int_{-1}^1dx\int_0^Ne^{-3y}|\tilde{\xi}_{\bf k}^{(0)}|^2dy.
\label{eq_VEVmassperturbative}
\end{equation}
The entire coupled dynamics is finally described by the following system of equations:
\begin{equation}
\begin{cases}
\frac{d^2\tilde{\xi}^{(0)}_{\bf k}}{dN^2}+\frac{d\tilde{\xi}_{\bf k}^{(0)}}{dN}+\left[e^{-2y}+\delta^2\tilde{\mathcal{A}}_i^2-2\right]\tilde{\xi}_{\bf k}^{(0)}=0\\
\frac{d^2\delta\tilde{\xi}_{\bf k}}{dN^2}\!+\!\frac{d\delta\tilde{\xi}_{\bf k}}{dN}\!+\!\left[e^{-2y}+\delta^2\tilde{\mathcal{A}}_i^2-2\right]\delta\tilde{\xi}_{\bf k}=-2e^{-y}x\delta\tilde{\mathcal{A}_i}\tilde{\xi}_{\bf k}^{(0)}\\
\begin{split}
\frac{d^2\tilde{\mathcal{A}}_i}{dN^2}\!&+\!\frac{d\tilde{\mathcal{A}}_i}{dN}\!+\!\left[\!\frac{\sigma^2}{2}\!\!\int_{-1}^1\!\!\!\!dx\!\int_0^N\!\!\!\!\!e^{-3y}|\tilde{\xi}_{\bf k}^{(0)}|^2dy\!-\!\alpha(\alpha\!+\!1)\right]\!\tilde{\mathcal{A}}_i=\\&=-\zeta\int_{-1}^1dx\int_{0}^Ne^{-4y}x\text{Re}\left\{\tilde{\xi}_{\bf k}^{(0)*}\delta\tilde{\xi}_{\bf k}\right\}dy
\end{split}
\end{cases}\!\!\!\!.
\label{eq_coupledFINALsystem}
\end{equation}
We remark that this approach is valid as long as the scalar modes $\tilde{\xi}_{\bf k}$ are almost constant with $x$. To check this, we define the ratio
\begin{equation}
R(N)\equiv \text{max}_{x\in[-1,1], y\in[0,N]}\left\{\frac{|2e^{-y}x\delta\mathcal{A}_i|}{|e^{-2y}+\delta^2\mathcal{A}_i^2|}\right\},
\label{eq_R}
\end{equation}
and the perturbative approach remains valid as long as $R(N)\ll 1$. Checking that $R(N)\ll 1$ is thus fundamental in order to guarantee the reliability of our work, and for this reason we will show this check for all our simulations.

\subsection{Reduction of the background electric energy density via the Schwinger effect}

Simulating the coupled dynamics, we can finally compute the resulting electric energy density, in presence of Schwinger particle production. In terms of the quantities used in the numerical simulation, it is written as
\begin{equation}
\frac{\rho_{\text{E}}^{(\text{in})}(N)}{M^4_{\text{Pl}}}=\frac{1}{2}\left(\frac{H_I}{M_{\text{Pl}}}\right)^4\left(\frac{d\tilde{\mathcal{A}_i}}{dN}-\alpha\tilde{\mathcal{A}_i}\right)^2e^{-2N}.
\label{eq_numericalrhoE}
\end{equation}
From a numerical viewpoint, initial conditions can be imposed setting $\tilde{\mathcal{A}_i}(0)$ and $\frac{d\tilde{\mathcal{A}}_i}{dN}(0)$. In particular, we determine them imposing that $\rho^{(\text{in})}_{\text{E}}(N=0)$ obtained for $e\neq 0$, from Eq. (\ref{eq_numericalrhoE}), is equal to $\rho^{(\text{in})}_{\text{E}}(\theta=1)$ obtained for $e=0$, from Eq. (\ref{eq_CLASSICALrhoE}). Specifically, we choose $\tilde{\mathcal{A}}_i(N=0)=0$ and determine $\frac{d\tilde{\mathcal{A}}_i}{dN}(0)$ according to the conditions above, getting $\frac{d\tilde{\mathcal{A}}_i}{dN}(0)\simeq 1.47$ for $\alpha=-2.5$.

Eq. (\ref{eq_numericalrhoE}), with the electromagnetic field computed numerically, is valid for $\theta>\theta_*$, before backreaction. In terms of $N$, the region $[\theta_*,1]$ corresponds to $[0,-\ln(\theta_*)]\simeq [0,14.33]$. However, we want to extend the domain of our solution for the whole duration of inflation. Accordingly:
\begin{itemize}
\item[-] we numerically solve the system with $\alpha=-2.5$ until $N_*=14.33$, imposing the initial conditions discussed above, and then
\item[-] we set new initial conditions at $N_*=14.33$, guaranteeing continuity and differentiability of the electric energy density at $N_*$, and solve the system with $\alpha=-2$ until the end of inflation.
\end{itemize}
In this way, we obtain the full curve of the background electric energy density during inflation when the Schwinger effect is considered.

In addition, we compute the energy density of the charged particles, say $\rho_{\chi}$. As the complex scalar fields are quantum fields, we compute their energy density from the VEV of the 00 component of the respective energy momentum tensor. In particular, we arbitrarily choose to show only the free component of the whole $T_{00}$ associated with the $\chi-$field, say $\rho_{\chi}^0$, without considering the interaction term with the electromagnetic field. In this way, we aim to compare on the one hand the free electric energy density, and on the other hand the free energy density associated with the produced particles. Indeed, we want to show that, as charged particles are produced via the Schwinger effect, the electric fields are drastically suppressed. In terms of energy, we would like to see that the energy density associated with electric fields is reduced as the energy density associated with the $\chi-$field alone becomes relevant with respect to the background.

\begin{figure*}
\includegraphics[width=1.\textwidth]{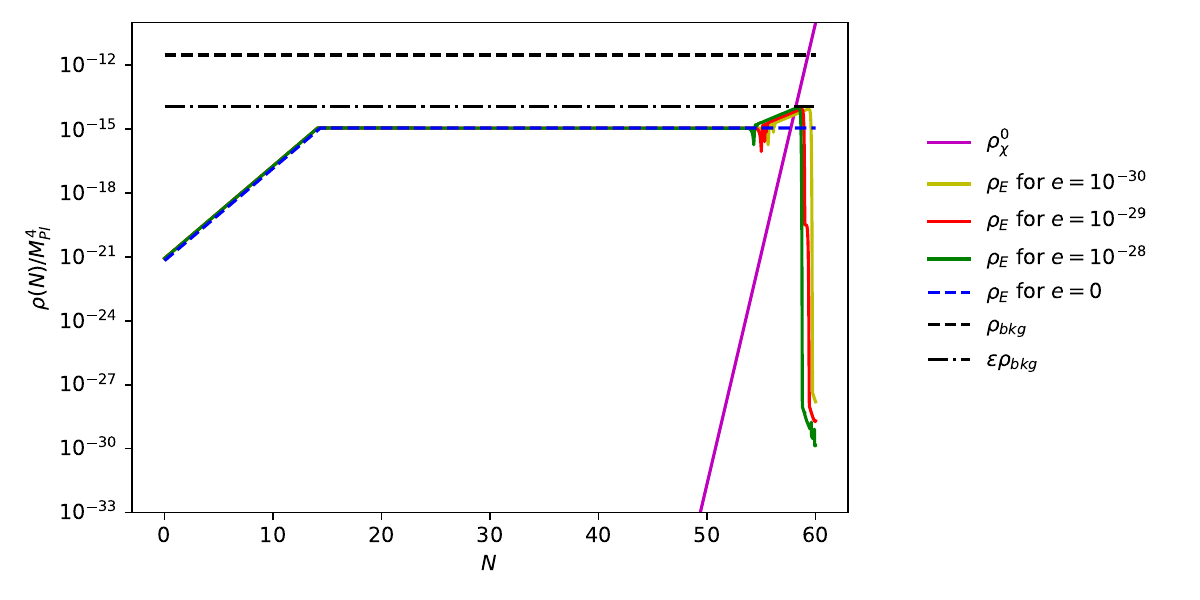}
\caption{The figure shows the electric energy density obtained with $e=0$ (dashed blue line) and $e\neq 0$ in comparison. In particular, we show the curves for $e=10^{-28}$ (green line), $e=10^{-29}$ (red line) and $e=10^{-30}$ (yellow line). In addition, we also show the energy density associated with the produced charged scalar particles (purple line).}
\label{fig_SchwingerPLOT}
\end{figure*}

Using the quantum expression for the scalar field in Eq. (\ref{eq_quantumchi}), we obtain
\begin{equation}
\begin{split}
\frac{\rho^0_{\chi}(N)}{M^4_{\text{Pl}}}&=\frac{1}{2\pi^2}\left(\frac{k_{\text{i}}}{M_{\text{Pl}}}\right)^2\left(\frac{H_I}{M_{\text{Pl}}}\right)^2e^{3N}\times \\&\times \int_0^Ne^{-3y}\left(\left|\frac{d\tilde{\xi}_{\bf k}^{(0)}}{dN}\right|^2+e^{-2y}\left|\tilde{\xi}_{\bf k}^{(0)}\right|^2\right)dy.
\end{split}
\label{eq_rhoCHI}
\end{equation}
We simulate the dynamics for three values of the coupling constant $e$, specifically $e=10^{-28}$, $e=10^{-29}$ and $e=10^{-30}$. In Fig. \ref{fig_SchwingerPLOT}, we show the curves of the electric energy density obtained for the three different values of $e$. In addition, we show the curve of the energy density of the produced scalar particles, that turns out to be the same for all the values of $e$.

We can see that, for most of the inflation, the curves obtained for $e=0$ and $e\neq 0$ coincide. In this phase, the charged particles produced via the Schwinger effect do not backreact on the electric background yet.

\begin{figure}
\includegraphics[width=0.5\textwidth]{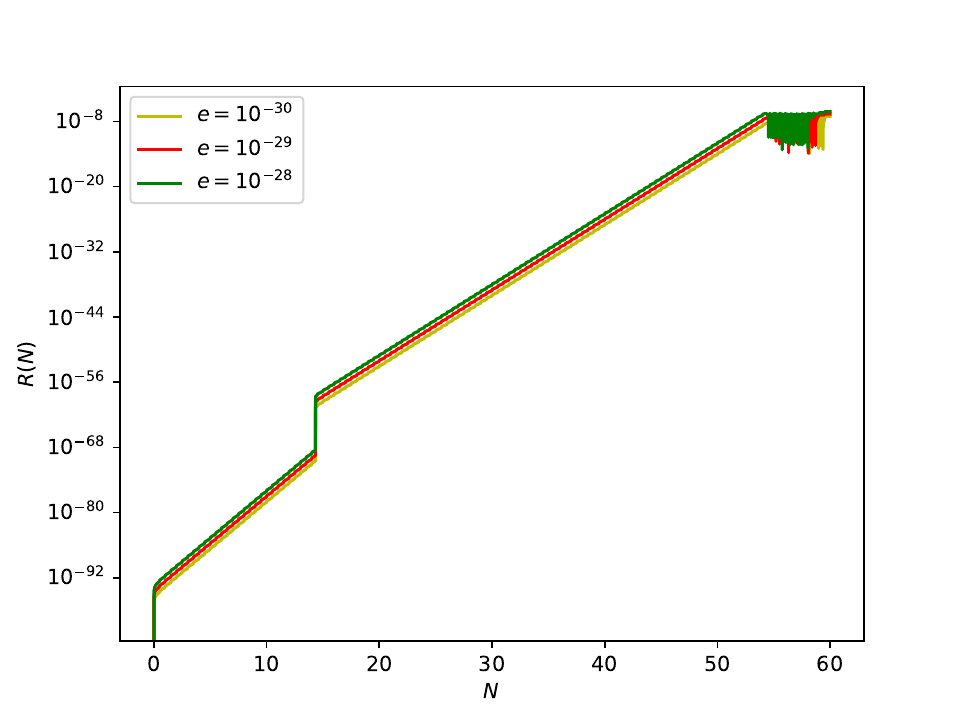}
\caption{We show the function $R(N)$ obtained for $e=10^{-28}$ (green line), $e=10^{-29}$ (red line) and $e=10^{-30}$ (yellow line).}
\label{fig_R}
\end{figure}

Then, once the charged particles have been produced in a significant amount, their backreaction affects the electric background. Accordingly, the curves obtained for $e\neq 0$ deviate significantly from the ones obtained with $e=0$. In particular, in this phase we can distinguish two different regimes.
\begin{itemize}
\item[-] Firstly, $\rho_{\text{E}}$ performs very frequent oscillations. In fact, for a better visibility of the plot, we show only the maxima of these oscillations. The resulting curves, initially smoothly increasing and then almost vertically decreasing, show that the average value of $\rho_{\text{E}}$ decreases during this phase.

Looking at the equation for $\tilde{\mathcal{A}}$, the oscillations start when the effective mass squared becomes positive, that is, when the term coming from $\braket{\xi\xi^*}$ overcomes the term due to the kinetic coupling with the inflaton field. When this occurs, the tachyonic regime that has previously driven the amplification of the electric fields gives way to a strongly oscillatory phase. In particular, as the contribution coming from $\braket{\xi\xi^*}$ continues to grow due to particle production, the electromagnetic field becomes heavier and heavier, and the increasing inertia makes the amplitude of the oscillations exponentially decay. In this phase, the source is still negligible with respect to the mass term, that accordingly drives the whole dynamics.
\item[-] When the source becomes comparable to the mass term, the dynamics enters an asymptotic regime. In this respect, we define the function
\begin{equation}
Q(N)\equiv \frac{\left|-\zeta\int_{-1}^1dx\int_0^Ne^{-4y}x\text{Re}\left\{\tilde{\xi}_{\bf k}^{(0)*}\delta\tilde{\xi}_{\bf k}\right\}dy\right|}{\left|\left[\frac{\sigma^2}{2}\int_{-1}^1dx\int_0^Ne^{-3y}\left|\tilde{\xi}_{\bf k}^{(0)*}\right|^2dy-\alpha(\alpha+1)\right]\tilde{\mathcal{A}_i}\right|},
\label{eq_Q}
\end{equation}
whose plot\footnote{The discontinuity in Fig. \ref{fig_Q} is due to our choice of abruptly changing from $\alpha=-2.5$ to $\alpha=-2$, that we have already discussed many times.} is shown in Fig. \ref{fig_Q} for all the values of $e$ considered in this work.

\begin{figure}
\includegraphics[width=0.5\textwidth]{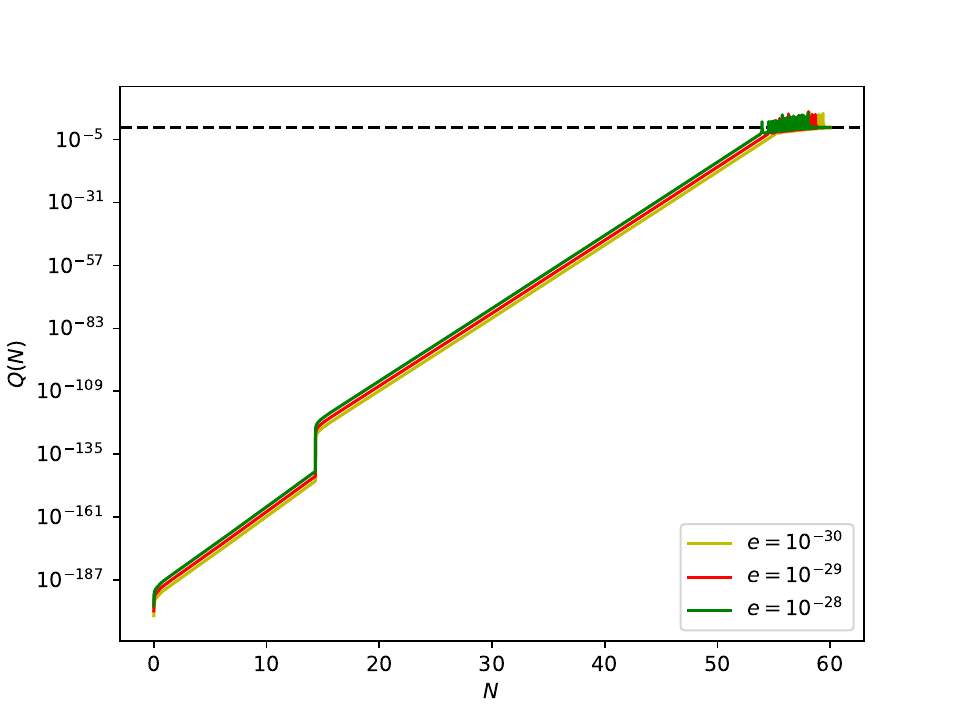}
\caption{We show the function $Q(N)$ obtained for $e=10^{-28}$ (green line), $e=10^{-29}$ (red line) and $e=10^{-30}$ (yellow line). The black dashed line is in correspondence of $Q(N)=1$.}
\label{fig_Q}
\end{figure}

We can check that the value of $Q$ grows as the inflation proceeds, approaching 1 in the final stages. This means that the source is negligible with respect to the mass term for most of the inflation, while at its end the two become comparable. In this phase, the equation gets rather simplified to
\begin{equation}
\frac{d^2\tilde{\mathcal{A}}_i}{dN^2}+\frac{d\tilde{\mathcal{A}}_i}{dN}\simeq 0.
\label{eq_EOM-A-asymptotic}
\end{equation}
Accordingly, the asymptotic solution for the electromagnetic field has a simple form,
\begin{equation}
\tilde{\mathcal{A}}_i^{(\text{asym})}(N)\simeq c_1+c_2e^{-N},
\label{eq_A-asymptotic}
\end{equation}
and in correspondence of that the asymptotic solution for the electric energy density reads
\begin{equation}
\frac{\rho_{\text{E}}^{(\text{asym})}(N)}{M^4_{\text{Pl}}}\simeq \frac{1}{2}\left(\frac{H_I}{M_{\text{Pl}}}\right)^4\left[\alpha c_1+c_2(1+\alpha)e^{-N}\right]^2e^{-2N}.
\label{eq_A-asymptotic}
\end{equation}
Eq. (\ref{eq_A-asymptotic}) shows an exponential decay of the electric energy density, that we can see in Fig. \ref{fig_SchwingerPLOT} in the very final stages of the inflation. 
\end{itemize}
In addition, for the sake of consistency, in Fig. \ref{fig_R} we show\footnote{As in Fig. \ref{fig_Q}, the discontinuity is due to the analytical approximation of instantaneously passing from $\alpha=-2.5$ to $\alpha=-2$.} the function $R(N)$ for $e=10^{-28}$, $e=10^{-29}$ and $e=10^{-30}$. We can check that, as anticipated, $R(N)\ll 1$ in all our simulations. Moreover, the maximum value of $R$ does not seem to depend on the value of $e$, making our perturbative approach generally valid.

Finally, two comments need to be made about Fig. \ref{fig_SchwingerPLOT}.
\begin{itemize}
\item[-] The energy density of the charged particles produced via the Schwinger effect becomes comparable to $\rho_{\text{bkg}}$ at a certain time, and then even greater. This means that, after that time, the de-Sitter approximation should not be valid anymore, and accordingly the entire treatment should be reviewed. 
\item[–] Due to the Schwinger particle production, at a certain point the electric energy density becomes lower than the magnetic and longitudinal components. Therefore, in principle, we can no more rely on the separation between background and perturbative dynamics adopted till now, and our approach should be accordingly reviewed.
\end{itemize}

However, addressing these two points would require a very expensive numerical computation, and the benefits of this kind of analysis would regard only the very final stages of inflation, or even later epochs. In addition, we remark that our approach would stop to be valid when the suppression of the electric fields due to the exponential Schwinger production is already clearly visible. Therefore, our main results remain valid, and eventual small deviations may occur only in the very final stages of inflation, or even later, once the effect of the Schwinger production has been already manifested. For these reasons, we do not include this treatment in our paper, and defer it to future works.

\section{Final outlooks and perspectives}{\label{conclusions}}

In this work, we investigated inflationary magnetogenesis with massive vector fields, including charged scalar particle production via the Schwinger effect and its induced backreaction on the electric background. By assigning a nonzero mass to the vector field, we explicitly break gauge invariance in the Maxwell Lagrangian, and the longitudinal modes therefore become physical. Additional contributions then arise in the total electromagnetic energy density, whose magnitude must be evaluated. To this end, we initially neglected the Schwinger effect and considered fully quantum electromagnetic fields. We derived the transverse and longitudinal mode equations and obtained their complete analytical solutions. As a result, the transverse modes remain essentially unaltered by the presence of a photon mass within experimental bounds, and accordingly they give rise to a dominant electric energy density and a perturbative magnetic component. In particular, we estimated the present magnetic field amplitude to be $B_0\simeq 2.67\times 10^{-17}\text{ G}$, fully consistent with observational constraints. On the other hand, longitudinal modes provide only perturbative contributions to the total electromagnetic energy density, comparable to the magnetic part. Therefore, in the massive framework, the dominant contribution is still given by the electric energy density, constituting the background, while magnetic and longitudinal components behave as perturbations. This result allows us to treat the electric background separately from the remaining contributions.

We then incorporated the Schwinger effect and investigated its impact only on the background electric energy density. Indeed, we showed that the electric component represents the dominant term in the total electromagnetic energy density, while Schwinger particle production constitutes a nonperturbative effect acting primarily on this dominant background. The magnetic and longitudinal energy densities remain unaffected by Schwinger production as long as they behave as perturbations with respect to the background. We addressed this problem using a classical electromagnetic field and a quantum complex scalar field interacting through a minimal coupling, thereby adopting a semiclassical approach. As a consistency check, we showed that for a null coupling constant the resulting electric energy density coincides with that obtained within the fully quantum framework. For non-null coupling constants, we numerically obtained the evolution of the electric energy density in the presence of the Schwinger effect. We found that electric fields are drastically suppressed in the final stages of inflation as a consequence of the exponential growth of the energy density of the produced scalar particles.

Future works could aim at constructing a fully gauge invariant formulation and investigating its implications for magnetic and electric field generation. In particular, gauge invariance could be restored within the Maxwell-Proca Lagrangian through the Stueckelberg mechanism by introducing a new real scalar field. The physical interpretation of such a scalar should also be discussed. Moreover, fermionic particle production via the Schwinger effect could be investigated instead of charged scalar production. It would also be fruitful to explore alternative methods for treating Schwinger production without necessarily separating the background evolution from perturbative contributions. Finally, it would be intriguing to unify the magnetogenesis and baryogenesis frameworks within a single, more general scenario capable of describing both electromagnetic field generation and the origin of the baryon asymmetry.

\section*{Acknowledgements}
MD and OL are grateful to Antonino Giacomo Marino for fruitful discussions on magnetogenesis. The authors acknowledge the support of the Istituto Nazionale di Fisica Nucleare (INFN) through the Iniziativa Specifica QGSKY. OL also acknowledges Ginestra Bianconi and the hospitality of Queen Mary University of London, where this work was finalized.

\bibliography{bibliography}

\appendix

\section{Useful functions}{\label{app_usefulfunctions}}

The functions $e_1(x)$ and $e_2(x)$ appearing within Eq. (\ref{eq_rhoEperpAFTER}) are defined as follows:
\begin{widetext}
\begin{subequations}
\begin{align}
&e_1(x)=H^{(1)}_{\alpha-\frac{1}{2}}(x)\left(Y_{-\frac{3}{2}}(x)-Y_{-\frac{7}{2}}(x)\right)-Y_{-\frac{5}{2}}(x)\left(H^{(1)}_{\alpha+\frac{1}{2}}(x)-H^{(1)}_{\alpha-\frac{3}{2}}(x)\right),
\label{eq_e1rhoEperp}\\
&e_2(x)=H^{(1)}_{\alpha-\frac{1}{2}}(x)\left(J_{-\frac{3}{2}}(x)-J_{-\frac{7}{2}}(x)\right)-J_{-\frac{5}{2}}(x)\left(H^{(1)}_{\alpha+\frac{1}{2}}(x)-H^{(1)}_{\alpha-\frac{3}{2}}(x)\right).
\label{eq_e2rhoEperp}
\end{align}
\end{subequations}
\end{widetext}
The functions $g_1(x)$ and $g_2(x)$ appearing within Eq. (\ref{eq_rhoBAFTER}) are defined as follows:
\begin{widetext}
\begin{subequations}
\begin{align}
&g_1(x)=H^{(1)}_{\alpha+\frac{1}{2}}(x)\left(Y_{-\frac{1}{2}}(x)-Y_{-\frac{5}{2}}(x)\right)-Y_{-\frac{3}{2}}(x)\left(H^{(1)}_{\alpha+\frac{3}{2}}(x)-H^{(1)}_{\alpha-\frac{1}{2}}(x)\right),
\label{eq_g1rhoB}\\
&g_2(x)=H^{(1)}_{\alpha+\frac{1}{2}}(x)\left(J_{-\frac{1}{2}}(x)-J_{-\frac{5}{2}}(x)\right)-J_{-\frac{3}{2}}(x)\left(H^{(1)}_{\alpha+\frac{3}{2}}(x)-H^{(1)}_{\alpha-\frac{1}{2}}(x)\right).
\label{eq_g2rhoB}
\end{align}
\end{subequations}
\end{widetext}
The functions $d_1(x)$ and $d_2(x)$ appearing within Eq. (\ref{eq_rhoLperpAFTER}) are defined as follows:
\begin{widetext}
\begin{subequations}
\begin{align}
&d_1(x)=\frac{H^{(1)}_{\alpha+\frac{1}{2}}(x)Y_{-\frac{5}{2}}(x)-H^{(1)}_{\alpha-\frac{1}{2}}(x)Y_{-\frac{3}{2}}(x)}{J_{-\frac{3}{2}}(x)Y_{-\frac{5}{2}}(x)-J_{-\frac{5}{2}}(x)Y_{-\frac{3}{2}}(x)},
\label{eq_d1rhoLperp}\\
&d_2(x)=\frac{J_{-\frac{3}{2}}(x)H^{(1)}_{\alpha-\frac{1}{2}}(x)-J_{-\frac{5}{2}}(x)H^{(1)}_{\alpha+\frac{1}{2}}(x)}{J_{-\frac{3}{2}}(x)Y_{-\frac{5}{2}}(x)-J_{-\frac{5}{2}}(x)Y_{-\frac{3}{2}}(x)}.
\label{eq_d2rhoLperp}
\end{align}
\end{subequations}
\end{widetext}

\end{document}